\documentclass[12pt]{article}
\usepackage{graphicx,amsmath,amsfonts}
\usepackage{setspace}
\usepackage{caption}
\usepackage{apacite}
\usepackage{url}
\usepackage{lineno}
\usepackage[titletoc,title]{appendix}%
\usepackage{comment}
\usepackage{tabularx}
\usepackage{tabu}
\usepackage{lipsum}
\usepackage{enumitem}
\usepackage{natbib}
\usepackage[T1]{fontenc}
\usepackage[table]{xcolor}
\usepackage{caption}
\usepackage{float}
\restylefloat{figure}

\renewcommand{\figurename}{Fig.}
\renewcommand{\tablename}{Table}
\newcommand{\sectionname}[1]{Section }

\title{Identifying Solar Flare Precursors Using Time Series of SDO/HMI Images and SHARP Parameters}

\author{Yang Chen\footnote{Email: ychenang@umich.edu}\and Ward B. Manchester\and 
     Alfred O. Hero\and
     Gabor Toth\and
     Benoit DuFumier\and Tian Zhou\and Xiantong Wang\and 
     Haonan Zhu\and Zeyu Sun \and
     Tamas I. Gombosi}
\date{University of Michigan, Ann Arbor}

\begin{document}

\maketitle

\begin{abstract}
In this paper we present several methods to identify precursors that show great promise for early predictions of solar flare events. A data pre-processing pipeline is built to extract useful data from multiple sources, Geostationary Operational Environmental Satellites (GOES) and Solar Dynamics Observatory (SDO)/Helioseismic and Magnetic Imager (HMI), to prepare inputs for machine learning algorithms. Two classification models are presented: classification of flares from quiet times for active regions and classification of strong versus weak flare events. We adopt deep learning algorithms to capture both spatial and temporal information from HMI magnetogram data. Effective feature extraction and feature selection with raw magnetogram data using deep learning and statistical algorithms enable us to train classification models to achieve almost as good performance as using active region parameters provided in HMI/Space-Weather HMI-Active Region Patch (SHARP) data files. {Case studies show a significant increase in the prediction score around 20 hours before strong solar flare events.} 
\end{abstract}

\section{Introduction}
\label{sec:intro}
 

Observations have established that solar eruptions are all associated with highly nonpotential magnetic fields that store the necessary free energy. The most energetic flares come from the intense kilogauss fields of Active Regions (ARs), where free energy is stored with field-aligned electric currents. Magnetic energy release occurs across an enormous range of scales from the most energetic flares ($10^{32-33}$ erg) associated with high-speed Corona Mass Ejections (CMEs) down to ever-present nano-flares possibly heating the quiet corona ($10^{22-24}$ erg). According to the \citet{NOAAscale}, {during solar cycle 24,} there were  $>2000$ M flares, while there were less than $180$ X flares. The complexity of solar flares and the infrequent occurrence of energetic events makes fast and accurate predictions of the time and intensity multiple hours/days ahead an extremely challenging task. What exacerbates the situation for data-driven methods is the computational cost required to process the high-resolution and high cadence observations over an extended period of time. In the last few years, predictions of flares with data-driven approaches are getting more attention. 



Machine learning algorithms have been applied to solar eruptions only some two decades after {ML algorithms} were used to investigate the terrestrial impacts of solar storms. Several teams \citep[]{Huang:2018a, Song:2009a, Yu:2009a, Yuan:2010a, Ahmed:2013a} have forecast solar flares by using machine learning algorithms trained with parameters calculated from maps of the line-of-sight (LOS) component of the photospheric magnetic field observed by the Michelson Doppler Imager (MDI) instrument aboard the SOHO (Solar \& Heliospheric Observatory) spacecraft. {\citet{boucheron2015prediction} adopt the support vector machine for time series classification with the MDI data from 2000 to 2010.} {However, these studies rely on proxies found to be correlated to the nonpotential magnetic fields with strong shear measured by vector magnetographs.} 

Studies followed which applied the full vector magnetic field observations.  \citet{barnes2007probabilistic} were the first to use vector magnetograms to investigate solar flare forecasting using a statistical classifier, which outperforms the NOAA's SWPC prediction results~\citep{Crown:2012a, jolliffe2012forecast}. \citet{bobra2015solar} followed this with the first solar flare forecast using machine learning algorithms trained with parameters calculated from vector magnetic fields observed with the Solar Dynamics Observatory (SDO)/Helioseismic and Magnetic Imager (HMI). The magnetic field maps used in this case are spatially restricted to the near proximity of ARs, so called Space-weather HMI Active Region Patches, or SHARPs \citep{Bobra:2014a}. {The FLARECAST framework, an automated forecasting system, (\url{http://flarecast.eu/}) was developed by a European consortium ~\citep{florios2018forecasting}.} 
\citet{nishizuka2018deep} developed a solar flare prediction model using a deep neural network (DNN). {Further}, \citet{hada2016deep} attempted the real time automated forecast of solar flares with deep learning approaches. For a comprehensive review, see \citet{Leka:2018b} and \citet{Camporeale:2019a}.


Solar flares show dynamic behavior observed in the chromosphere, transition region and low corona \citep{Benz:2016a} that many studies have shown have a statistical correlation with flare production. These observations provide significantly more data for building a predictive model that uses images made across multiple wavelengths. \citet{Nishizuka:2017a} were the first to use machine learning algorithms to predict solar flares by not only parameterizing photospheric magnetograms but also using images of the chromosphere. Finally, \citet{Jonas:2018a} were the first to predict solar flares by using a machine learning algorithm along with maps of the photosphere, chromosphere, transition region, and corona, which is comparable in performance with the models of \citet{bobra2015solar} and \citet{Nishizuka:2017a}. 

In this paper, we {discuss the performances of our adopted} machine learning algorithms for time series classification and feature extraction based on image reconstruction, using HMI/SHARP patches and GOES data from {May 1, 2010} to {June 20, 2018}, toward {encouraging} solar flare (predictive) classifications. 
We use a Long Short Term Memory (LSTM) model~\citep{hochreiter1997long,gers1999learning} to classify solar flare events (B/C/M/X class) versus non-flare and strong flare (M/X class) versus weak flare (B class) using SHARP parameters several hours/days prior to the start or time of  peak intensity of the event. These SHARP parameters may be thought of as handcrafted features in machine learning in that they are selected based on physical understanding of quantities related to flare production {(see \citet{Bobra:2014a} and references therein; \citet{Leka:2003a} and references therein)}. In this case, they include a hierarchy of quantities characterizing the observed magnetic field such as magnetic flux, electric currents and current helicity. The LSTM model predicts binary outcomes using trained nonlinear transformations of input parameters and is shown to work for accurate classifications for time-series data~\citep{Goodfellow-et-al-2016}, including natural language text compression and speech recognition~\citep{graves2009novel,graves2013speech}. It should be noted that in the {majority of} previous work, static features are used for predictions, whereas in this paper we use time series for predictions and account for time-dependency instead of simply stacking up features from multiple time points {and ignoring the sequential nature of the features}, as is done in ~\citet{boucheron2015prediction} and~\citet{leka2018nwra}. {Features from multiple time points, when vectorized, are typically regarded as ``independent'' or ``pairwise dependent'' features/dimensions by most machine learning algorithms; whereas time series of features preserve the temporal structure, which could possibly be learned by appropriately training machine learning algorithms.} We then perform binary classification of strong/weak flares, replacing the SHARP parameters with machine-learned features. This includes three steps: 
\begin{enumerate}
    \item We derive features from vector magnetogram maps using the autoencoder, a deep learning technique that derives essential features to reconstruct images;
    \item We apply the marginal screening technique to remove redundant features for solar flare classification, which turns out to help avoid over-fitting effectively; and
    \item We train the LSTM model using the remaining features for classifications.
\end{enumerate} Our approach incurs differences in data preparation for machine learning tasks such that our results are not directly comparable with some examples in the literature {(see~\citet{jolliffe2012forecast,barnes2016comparison} for discussions on validation science).} 

{The remainder of the paper is organized as follows. We describe our general methodology in Section~\ref{sec:method}: including descriptions of the machine learning algorithms, data processing pipeline and data preparation for machine learning tasks, and evaluation metrics. In Section~\ref{sec:results}, we present our results for flare classifications, with SHARP parameters, and with machine-derived features; and we illustrate the flare classification models with several case studies. We conclude the paper in Section~\ref{sec:conclusion} with discussions of our promising results and future work.}

\tabulinesep=1ex
\newcommand\T{\rule{0pt}{2.5ex}} 
\newcommand\B{\rule[-1.2ex]{0pt}{0pt}} 
\rowcolors{2}{blue!9}{blue!3}

\section{Methodology}
\label{sec:method}
We provide a detailed description of the data pre-processing pipeline in \sectionname~\ref{subsec:pipeline}, while data preparation in the form of various training/testing sample splitting routines are discussed in \sectionname~\ref{subsec:data_split},  positive and negative classes are defined in \sectionname~\ref{subsec:pos_neg_prepare}, and metrics for evaluating different machine learning algorithms are given in Section \ref{subsec:metrics}, respectively. Finally, \sectionname~\ref{subsec:machine_learning_intro} gives a brief introduction to machine learning. 

\subsection{Data Pre-processing Pipeline}
\label{subsec:pipeline}

Our models use a time series of flare events from the NOAA Geostationary Operational Environmental Satellites (GOES) flare list \citep{Garcia:1994}. Classification is used for predicting discrete responses such as no flare (``quiet time'' of an AR), any flare (B/C/M/X class), weak flare (B class) or strong flare (M/X class). We use GOES data observed from 2010-05-01 to 2018-06-20 \citep{Garcia:1994} over which time there are $12,012$ solar flares listed with class, start, end, and peak intensity time of each event. {Flares of A class are omitted because their energy level is so low that they are frequently below the background brightness of the AR and consequently not counted in the GOES catalog.  The same is true of many B flares. If all were counted, the number of B flares would certainly outnumber the C flares. }

The flare events are then matched to the SHARP vector field data patches provided by the Joint Science Operations Center (JSOC) website. While the GOES flares are identified strictly with NOAA ARs, the SHARP patches are designed to include complete ARs and sets of ARs, so frequently a single HARP has multiple ARs, but it is unexpected that a single AR is split between HARPs (Todd Hoeksema, private communication). Our examination shows that $20\%$ of SHARP patches include components from multiple ARs.  This leads to a potential error where we may miss flare events occurring from within the SHARP but are attributed to a minor AR that was not counted. In the future, we will address the multiple-ARs-one-HARP problem by cutting the HARP regions into multiple ARs manually and then recalculating the SHARP parameters for each AR. 

The SHARP patches contain 2-D photospheric maps of $3$ orthogonal magnetic field components {observed with 1.0 arcsecond spatial resolution (0.5 arcsecond pixel size)} and provided with a time cadence of 12 minutes \citep{Hoeksema:2014a,Bobra:2014a}. From these data, parameters are calculated to specifically capture the structure and complexity of the magnetic field. As discussed in \citet{Leka:2003} and \citet{Bobra:2014a}, the parameters are designed to assess the flaring potential of ARs and are thus strongly representative of the total free energy of the magnetic field.  The free energy, in turn, is related to the electric currents flowing through the photosphere into the corona, which are proportional to the curl of the field $(\nabla \times \bf{B})$. These whole-active-region magnetic quantities can be effectively used as predictors of flares and also CMEs \citep[cf.][]{Falconer:2001, Falconer:2002, Falconer:2003, Leka:2003, Falconer:2006, Schrijver:2007, bobra2015solar}.  The SHARP parameters that we use are listed  in~\tablename~\ref{table:param} and further described in \citet{Bobra:2014a}. In addition, we also use NPIX, the number of pixels in a SHARP image, as a parameter.


%
\tabulinesep 0.5ex
\begin{table}[htb]
\centering
\caption{List of SHARP parameters and brief descriptions.}
\begin{tabu}{|>{\raggedright}p{1.0in}|>{\raggedright}p{3.75in}|}
\hline
\rowcolor{lightgray} \textbf{Parameter} \T\B & \textbf{Description} 
\\
\hline
TOTUSJH: \T& Total unsigned current helicity  \\
TOTUSJZ: & Total unsigned vertical current \\
SAVNCPP: & Sum of the modulus of the net current per polarity \\
USFLUX: & Total unsigned flux \\
ABSNJZH: & Absolute value of the net current helicity \\
TOTPOT: & Proxy for total photospheric magnetic free energy density \\
SIZE ACR: & De-projected area of active pixels ($B_z ?$ magnitude larger than noise threshold) on image in micro-hemisphere (defined as one millionth of half the surface of the Sun) \\
NACR: & The number of strong LoS magnetic-field pixels in the patch \\
MEANPOT: & Proxy for mean photospheric excess magnetic energy density \\
SIZE: & Projected area of the image in micro-hemispheres \\
MEANJZH: & Current helicity ($B_z$ contribution) \\
SHRGT45: & Fraction of area with shear $> 45^{\circ}$ \\
MEANSHR: & Mean shear angle \\
MEANJZD: & Vertical current density \\
MEANALP: & Characteristic twist parameter, $\alpha$  \\
MEANGBT: & Horizontal gradient of total field \\
MEANGAM: & Mean angle of field from radial \\
MEANGBZ: & Horizontal gradient of vertical field \\
MEANGBH: & Horizontal gradient of horizontal field \\
\hline
\end{tabu}
\vspace*{0.25em}
\label{table:param}
\end{table}
%

\begin{figure}[tbph]
    \centering
    \includegraphics[width=\textwidth]{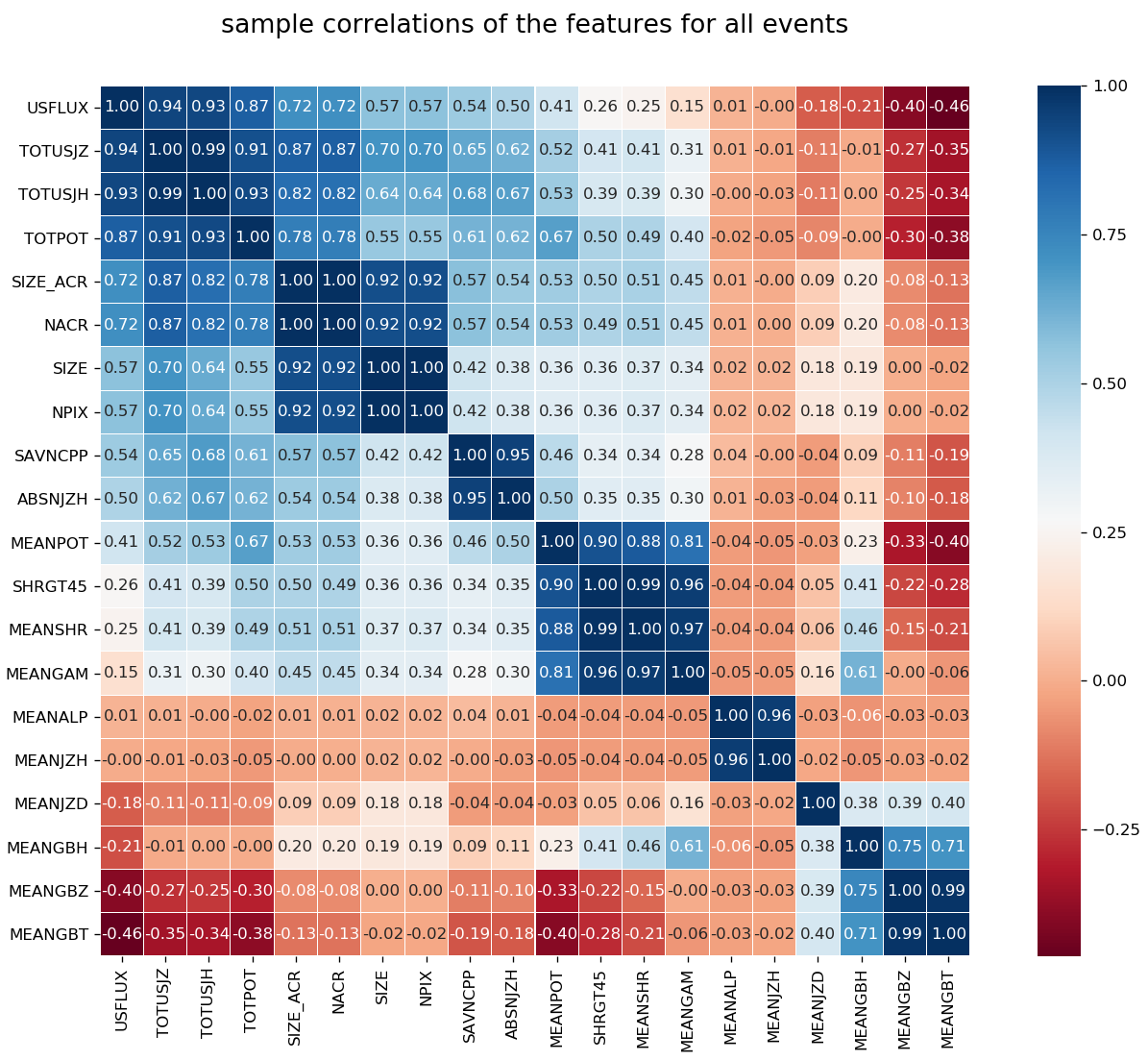}
\caption{Sample correlations of the features from all flare events that we consider. }
    \label{fig:corr_parameters}
\end{figure}

We recognize that these SHARP parameters are correlated with each other, in fact, some are highly correlated (even repetitive). \figurename~\ref{fig:corr_parameters} gives the sample correlations of these features from all B/C/M/X flares. In a PCA (principal component analysis, \citet{pearson1901liii}) study, we find that {the first} 7 principal components (linear combinations of these features) explain more than 95\% of the variability of the 20 features. Therefore, we do obtain an efficient dimension reduction via the PCA study: Using these 7 principal component is good enough for the subsequent machine learning task as opposed to the original 20 features. We have compared the performance of the machine learning tasks using all original 20 features as opposed to using these 7 principal components in Sections~\ref{subsec:strong_weak_classification} and~\ref{subsec:case_study}. Note that this is important to recognize because highly correlated (or redundant) features might cause various problems in the machine learning algorithm, such as non-identifiability and overfitting, both of which are results of the machine being ``confused'' about two almost identical variables, especially when evaluating which one is more important (a notion called variable importance in the machine learning literature, which we will talk about in \sectionname~\ref{subsec:variable_importance}). {This is a common problem in machine learning and is also acknowledged in previous studies of solar flare predictions, see e.g. \citet{Leka:2003,bobra2015solar,florios2018forecasting,tang2014feature,tolocsi2011classification} for more discussions.} 


{We built a data preparation pipeline that identifies SDO/HMI SHARP patches associated with solar flare events at any specified level as recorded in the GOES data set, and downloads the SHARP data files including the 3-component magnetogram data and the SHARP parameters for a specified number of hours prior to a solar flare event. The four steps are described as follows.}
\begin{enumerate}
    \item We first set a time range and download the whole GOES X-ray flare record. The queried items are: class and strength, NOAA AR number, event date, and the start, peak intensity, and end times of flare events.
    \item For each record in the GOES data set, we query the JSOC for the SHARP data with the end time equal to the flare peak intensity and decide the start time of the query based on how many frames we need with a 1 hour cadence. 
    \item We use the NOAA AR number in the GOES data set, provided 3 criteria are satisfied: (1) the NOAA number {in the HARP record} is the same as that in the GOES record; (2) the location of the AR is within $\pm 68 \deg$ from the central meridian {(in order to avoid projection effects \citep{bobra2015solar})}; (3) the time is before the peak intensity time.
    \item Finally, we download the data from JSOC based on SHARP number, cadence and the specified time frame.
\end{enumerate}


The data pre-processing pipeline described above gives us the list of flare events (of B/C/M/X classes), together with {the time series of features (SHARP parameters) and the magnetic images}. Now we describe how we feed these values into machine learning algorithms and on what the performance metrics are based.

\subsection{Details on Data Preparation: Training/Testing Splitting}
\label{subsec:data_split}

In order to properly calibrate the performance of the machine learning algorithms, we need to split the samples (flare events) into a training set and a testing set. The training data is used to train the machine learning models; and the testing data, which does not overlap with the training data, serves the purpose of calibrating the out-of-sample performance of the machine learning algorithms. We consistently take the ratio of training and testing samples to be $2:1$ for all models presented throughout the paper. 

Our default choice is the \textbf{Random-Splitting} scheme, which randomly selects flare events in the training and testing data. We run the random splitting 20 times for each model to guarantee the robustness and consistency of the results. This scheme does not take into account which AR a flare event is from, nor the year in which a flare event happened. Therefore, we also explore and test out other possible training/testing splitting methods: split-by-year and split-by-active-region. \tablename~\ref{tab:strong_weak_flares_years} lists the number of {flares of each class, i.e. B/C/M/X flares from 1100 ARs, recorded by the GOES data set corresponding to each year 2010 to 2018.} {Among the 1100 ARs that we process based on the GOES data set, the minimum number of flare events is 1 per AR and the maximum number of flare events is 141 per AR (given by AR 12297); 208 of the 1100 ARs have a strong flare (M/X class) associated.} The results of all the alternative training/testing splitting methods, which we summarize in Appendix~\ref{subsec:more_results}, turn out to be similar to the results based on random splitting we present in \sectionname~\ref{subsec:strong_weak_classification} for strong/weak flare classification and \sectionname~\ref{subsec:case_study} for case studies.


\begin{table}[b]
\centering
\caption{{
The number of flares of B/C/M/X class recorded in each year from 2010-05-01 to 2018-06-20 in the GOES data set. }}
    \begin{tabu}{|c|cccccccccc|}
    \hline
    \rowcolor{lightgray} Class/Year & 2010 & 2011 & 2012 & 2013 & 2014 & 2015 & 2016 & 2017 & 2018 & Total\\
    \hline
    X&0	&9	&7	&12	&16	&2	&0	&4	&0 & 50\\
M &0 &106 &124 &97 &194 &128 &15 &37 &0 & 701\\
C &1 &1002 &1115 &1197 & 1626 & 1275 &294 &229 & 11 & 6750\\
B &19 &665 &475 &469 & 184 & 446 & 757 & 620 & 207 & 3842\\
\hline
    \end{tabu}
    \label{tab:strong_weak_flares_years}
\end{table}

We test out two different sample splitting strategies based on \textbf{Split-by-Year}. (1) We randomly select several years' samples as the test set with the guarantee that the test samples are around 66\% of all the samples. {(2) We train with data from solar cycle 24, from years 2010-2013, when the sunspot activities see an increase and stabilize at maximum; and test on data from years 2014-2018, when the sunspot activities see a decrease.} 
We test out several different configurations based on \textbf{Split-by-Active-Region}. Prior to the splitting of test and training, we conduct a normalization step, which is designed to examine whether the model training is dominated by any particularly active-flaring AR. This is done by randomly selecting a limited number (which we call a ``cap'') of flares from each AR. The cap is set to be 2,3,4,5,10,15, and infinity (when we consider all flares). \tablename~\ref{tab:num_AR_num_events} gives the total number of ARs that have $1, 2, 3, 4, 5$, or $>5$ strong or weak flare events that we consider, {which is from the GOES data set. We note that here the number of B flares is under-recorded in the GOES data set, which is due to the fact that the B flares are not recorded when the ARs sustained emission exceeds the level of B flares.}
The number of ARs with a large number of flare events is not many, thus the possibility of flares from a single AR dominating the inference is not likely. Nevertheless, we test out our classification model with different ``cap'' numbers to rule out that possibility. We randomly select 67\% of the ARs (635 in total) as the ``training ARs'' and the remaining 33\% of the ARs as the ``testing ARs''. All observations for a chosen AR (with a maximum number of flare events bounded by the cap) are put either in the training or testing set, based on whether the AR is a ``training AR'' or a ``testing AR''. {See Appendix~\ref{subsec:more_results} for detailed results for both Split-by-Year and Split-by-Active-Region.}  

\begin{table}[tbph]
    \centering
    \caption{Number of ARs (ARs) corresponding to the specified number (1, 2, 3, 4, 5, and $>5$) of weak (B) and strong (M/X) flare events for each AR {recorded in the GOES data set}. }
    \begin{tabu}{|c|cccccc|}
    \hline
    \rowcolor{lightgray} Number of M/X Flares & 1 & 2 & 3 & 4 & 5 & $> 5$\\
    \hline
    Number of ARs & 60 & 31 & 13 & 10 & 7 & 29\\
    Number of B Flares & 1 & 2 & 3 & 4 & 5 & $> 5$\\
    Number of ARs & 321 & 148 & 51 & 19 & 2 & 0\\
    \hline
    \end{tabu}
    \label{tab:num_AR_num_events}
\end{table}

Furthermore, we normalize the data by subtracting the mean and dividing by the standard deviation {of the training data}, which is the most commonly adopted normalization method in practice~\citep[Section~7.10]{trevor2009elements}, before training the machine learning algorithms. We apply the same normalization to the testing data. Since the inputs of our machine learning algorithms are time series of SHARP parameters, we perform a global normalization of the whole time series of each feature: so as not to lose information in the normalization step. 

\subsection{Details on Data Preparation: Defining Positive/Negative Class}
\label{subsec:pos_neg_prepare}

In a binary classification task, such as strong/weak flare classification, to give sensible results, we need to prepare the data by defining the positive class (e.g. strong flares of M/X class) and negative class (e.g. weak flares of B class) properly to train and test the machine learning algorithm. Different preparations of positive and negative class could lead to different results (in terms of the metrics defined in \sectionname~\ref{subsec:metrics}), thus it is important to describe clearly what is done in this step. This is also the crucial step that makes different machine learning results noncomparable: if two researchers choose disparate positive/negative class preparations, the corresponding results cannot be compared fairly. {Clearly stating the data preparation, such as sample selection, for each machine learning tasks is a key step toward reproducibility of our results.} 

In our strong/weak flare classification models, we feed time series of features, for both the positive class (strong flares of M/X class) and negative class (weak flares of B class), into the machine learning algorithms. Therefore, it is important that the time series do not overlap significantly: otherwise, the features from the overlapping time points appear both in the positive and negative class, making it harder for the machine to differentiate. Besides, the forecasting window matters. For example, when we train a model to predict $72$ hours ahead of an M/X flare, if a B flare happens within this $72$ hour window, then the precursors that the machine could possibly find are predictive for both the M/X flares and B flares. Therefore, in our preparation of the positive and negative classes for the machine learning algorithms, we need to take all of these situations into account.  Intuitively, the longer the time series we use, and/or the longer the forecasting time, the more stringent the condition for selecting the positive and negative classes becomes. 
We will elaborate this again for strong/weak flare classifications and case studies in \sectionname~\ref{sec:results}. To make the results transparent and reproducible, we list the number of flare events of each class we use for training and testing the machine learning algorithms in \sectionname~\ref{sec:results} when we present our results. 

\subsection{Evaluation Metrics for Classification Algorithms}
\label{subsec:metrics}

Given that solar flare events, especially the intense ones, are relatively ``rare'', i.e. the ``positive class'' (a solar flare event)  is much smaller than the ``negative class'' (no solar flare event), we need evaluation metrics to quantify how well our models fit both the ``positive class'' and the ``negative class''. 
We use the following four metrics to evaluate our binary classifiers: the $F_1$ score, which is the harmonic mean of Precision and Recall, with the best value at $1$ and worst at $0$; the true skill statistic (TSS); and the Heidke skill scores ($\text{HSS}_1$ and $\text{HSS}_2$). See~\citet{bobra2015solar} for definitions of $\text{HSS}_1$ and $\text{HSS}_2$. We note that in the space weather community $\text{HSS}_2$ is referred to as the Heidke skill score \citep[cf.][]{Pulkkinen:2013a}. The higher the metrics (i.e. closer to $1$), the better the classifier. See~\citet{florios2018forecasting} for detailed descriptions for these skill scores. {Visually, we use the ROC (receiver operating characteristic) curves and the AUC (area under the ROC curves) values to examine the performances of the binary classifications presented in this paper \cite[see][for an introduction to ROC analysis]{fawcett2006introduction}.}

In the binary classification models, the raw output is a classification score that takes values between 0 and 1. This value represents the probability of the correct answer being positive (e.g. a strong flare in the strong/weak flare classification). We choose a default threshold, 0.5, for determining the predicted outcome. For example, we assign a predicted strong flare if the classification score is above 0.5 and a predicted weak flare if the classification score is below 0.5, in the strong/weak flare classification model. 


\subsection{Machine Learning and Statistical Algorithms}
\label{subsec:machine_learning_intro}

We give a brief introduction to the deep learning algorithms that we use to perform automatic feature extraction from HMI magnetograms (autoencoder for image reconstruction, marginal screening for feature selection) and solar flare classifications for time series observations (long short term memory networks).

%
 
Long Short Term Memory (LSTM) networks have been an effective solution to a wide range of ``sequence prediction problems'' such as image captioning, language translation, and handwriting recognition~\citep{graves2009novel,graves2013speech}. The LSTM network is a special kind of Recurrent Neural Networks (RNN) and it was first introduced by~\citet{hochreiter1997long} and improved by~\citet{gers1999learning}. It has internal contextual state cells that serve as memory cells, enabling information to flow from one step to the next. Thus, LSTM is capable of handling both short- and long-term dependencies. The LSTM network learns when to remember and when to forget through their forget gate weights. Consequently, the time dependency, whether short- or long-term, is also learned through the training of the algorithm. 

The autoencoder~\citep{liou2014autoencoder,kingma2013auto} neural network is an unsupervised learning algorithm that applies back propagation to learn structures of the input data such that the input and output are almost identical. The autoencoder network consists of the encoder, which transforms the input to ``code'', i.e. features, and the decoder, which transforms the ``code'' to the output~\citep[Chapter 14]{Goodfellow-et-al-2016}. The autoencoder is applied in our context to derive a relatively low-dimensional (vector) representation of the magnetogram field images (HMI images). 

%
%
Recall that our final objective is not magnetogram field image reconstruction. Instead, we are interested in classification: classifying large solar flare events versus weak/none solar flare events using features extracted from images. Therefore, we perform marginal screening to get rid of redundant features, which incurs over-fitting (i.e. worse performance), for the classification purpose \cite[see][for similar ideas applied to other models, including regression models]{tibshirani2003class,fan2008sure,fan2009ultrahigh,zhao2017marginal}. This method is typically used for genetic studies where thousands of genes (features) are considered for a disease/no-disease outcome whereas only a few genes are relevant for predicting the disease status, see e.g. the example in~\citet{hong2016data}. The marginal screening procedure goes as follows: we take one feature at a time and perform a two-sample $t$-test for testing the significance of the feature with respect to the binary outcome (e.g. strong versus weak flare); if the test turns out to be significant, we keep the feature; otherwise, the feature is deleted. We choose the significance value ($p$-value threshold) based on cross-validation of the classification results in the training data.

{
On the machine learning part, {our approaches enjoy the following nice properties.} 
\begin{enumerate}
    \item We perform feature extraction directly from HMI images using the deep learning algorithm autoencoder, as opposed to calculating various physical quantities from the observed AR magnetic field.
    \item We perform classification-oriented feature selection based on marginal screening, which effectively avoids over-fitting with a large number of features extracted.
    \item In our classification model, we adopt the LSTM, {which is also used in~\citet{hada2016deep}, that inputs time series data.} This takes into account the time evolution information instead of stationary features widely used in the literature for solar flare classifications as described in~\sectionname~\ref{sec:intro}.
    \item We compare the performance of the classification models using machine extracted features with those trained using SHARP parameters, which shows that potentially we could derive new features with machine-learning algorithms yet to be captured by well-known physical quantities (SHARP parameters). 
    \item We demonstrate the effectiveness and great potential of the proposed methods for early identification of precursors for strong flares by studying \textit{out-of-sample} prediction performances of trained models on four representative ARs.
\end{enumerate}
}

\section{Results} 
\label{sec:results}

We give the results of the solar flare classifications in this section. \sectionname~\ref{subsec:flare_no_flare} gives the results for the binary classification of ``solar flare events of any class'' against ``no solar flare events.'' {We also include a strong flare versus no flare classification, as in ~\citet{bobra2015solar}, in \sectionname~\ref{subsec:flare_no_flare}. We present the classification of strong and weak flares using SHARP parameters in \sectionname~\ref{subsec:strong_weak_classification}, discuss the feature importance in \sectionname~\ref{subsec:variable_importance}, and then use features learned directly from HMI magnetogram images in \sectionname~\ref{subsec:strong_weak_machine_features}. Case studies of strong/weak flare classification are given in \sectionname~\ref{subsec:case_study}.}

\subsection{Flare/Non-Flare Classification with SHARP Parameters}
\label{subsec:flare_no_flare}

\begin{figure}[htp]
    \centering
    \includegraphics[width=0.99\textwidth]{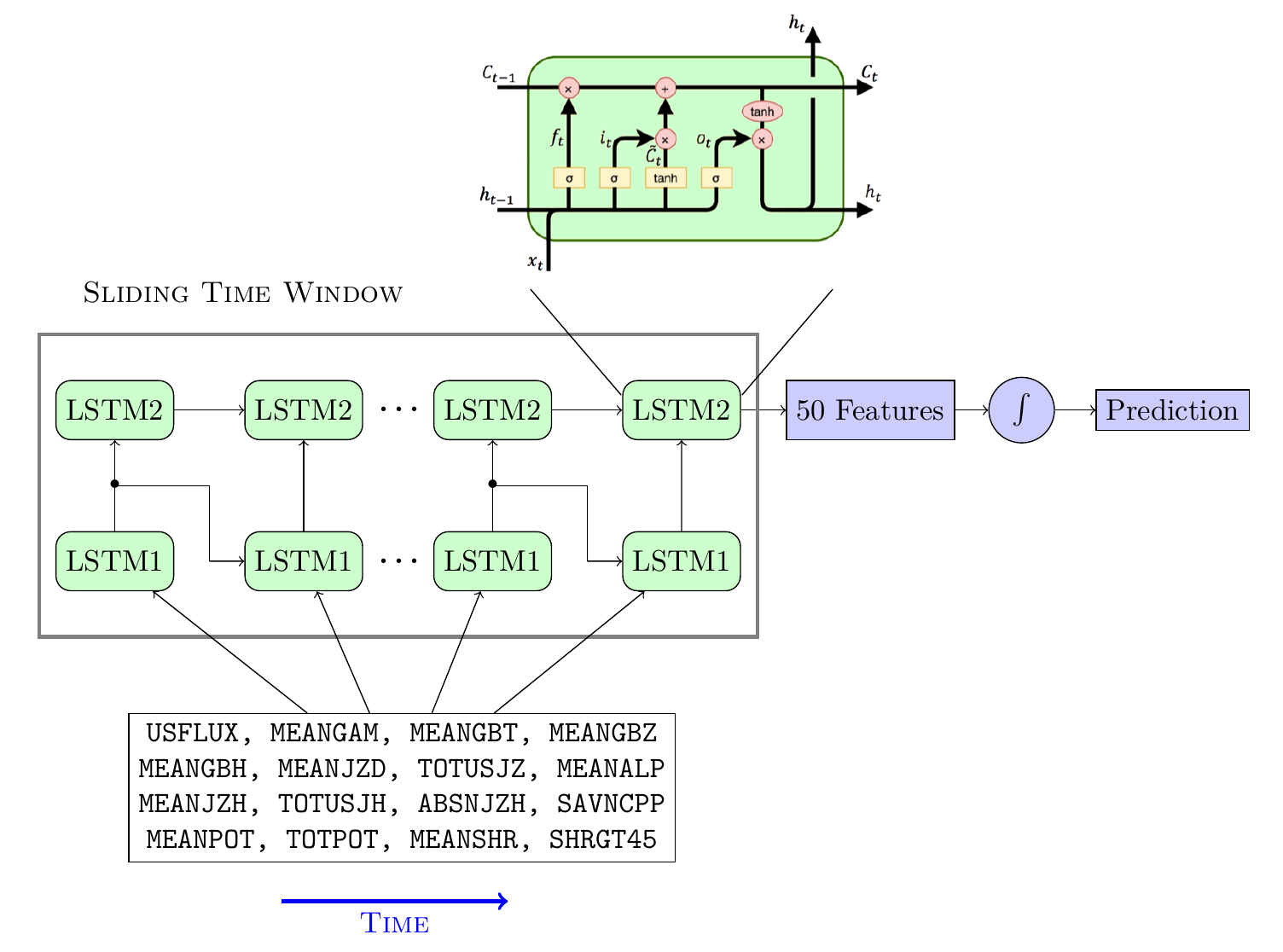}
    \caption{Flowchart of LSTM for classification using SHARP parameters from HMI/SDO header file (some are listed in the box at the bottom). These features can be replaced by other features, e.g. machined-learned features, see \sectionname~\ref{subsec:strong_weak_machine_features}.}
    \label{fig:flow_LSTM}
\end{figure}

We train an LSTM model for classifying flares of any intensity (positive class) against non-flares (negative class), using $20$ SHARP parameters (listed in \sectionname~\ref{subsec:pipeline}) at {1/3/6/12/24/48} hours preceding a solar flare event, at $1$ hour cadence. {\figurename~\ref{fig:flow_LSTM} shows a flowchart of LSTM for classifications with SHARP parameters. As reflected in \figurename~\ref{fig:flow_LSTM}, there are two LSTM layers, each of which contains a set of recurrently connected memory blocks. For each of the memory blocks (the green `LSTM1' or `LSTM2' boxes in \figurename~\ref{fig:flow_LSTM}), it takes the current input $x_t$, previous output $h_{t-1}$, and previous memory $c_{t-1}$, and generates a new output $h_t$ and memory $c_t$; see the detailed depiction of a memory block at the top of \figurename~\ref{fig:flow_LSTM}. Finally, since we are dealing with a binary classification problem, we adopt the sigmoid activation function as a dense output layer (the right purple blocks in \figurename~\ref{fig:flow_LSTM}).} The positive class consists of any solar flare (B/C/M/X) from the {239 HARP regions}. The members of the negative class are randomly selected to make sure that no flare event happens within $\pm 48$ hours. 
{After this selection, we will take into account around 100 ARs with around 200 flare/non-flare events for each forecasting window, which denotes the number of hours before the flare event (for the accurate numbers please see \tablename~\ref{table:no_events_first_flare}).} Note that the flares are rare and there are too many ``non-flares''. We randomly choose a subset of the non-flares to match the number of flares for training and testing.

\tabulinesep 0.5ex
\begin{table}[htb]
\centering
    \caption{The numbers of flares, non-flares, and ARs for each forecasting window (in hours, given in the first row) for M/X flare predictive classification model.}
\begin{tabu}{|c|ccccccc|}
    \hline
    \rowcolor{lightgray} Forecasting Window	& 1h	& 3h	& 6h	& 12h	&24h	& 48h	&72h\\
    \hline
    Number of Flares	&259	&259	&253	&250	&244	&206	&176\\
    Number of Non-Flares	&259	&259	&253	&250	&244	&206	&176\\
    Number of ARs	&122	&122	&119	&117	&112	&91	&81\\
    \hline
    \end{tabu}
    \label{table:no_events_first_flare}
\end{table}


%

\begin{figure}[b!]
\vspace*{1em}
\centering
\includegraphics[width=0.49\textwidth]{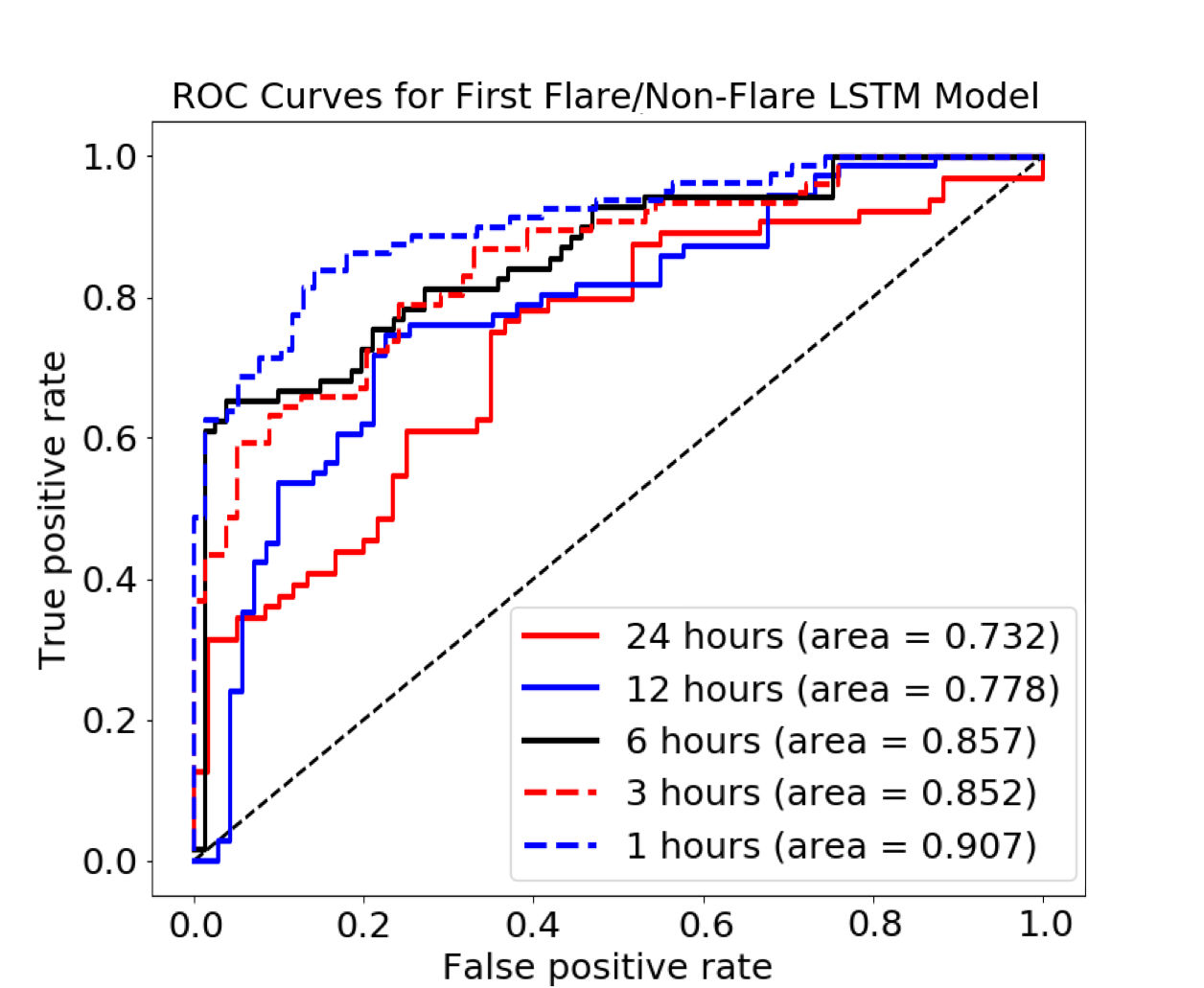}
\includegraphics[width=0.49\textwidth]{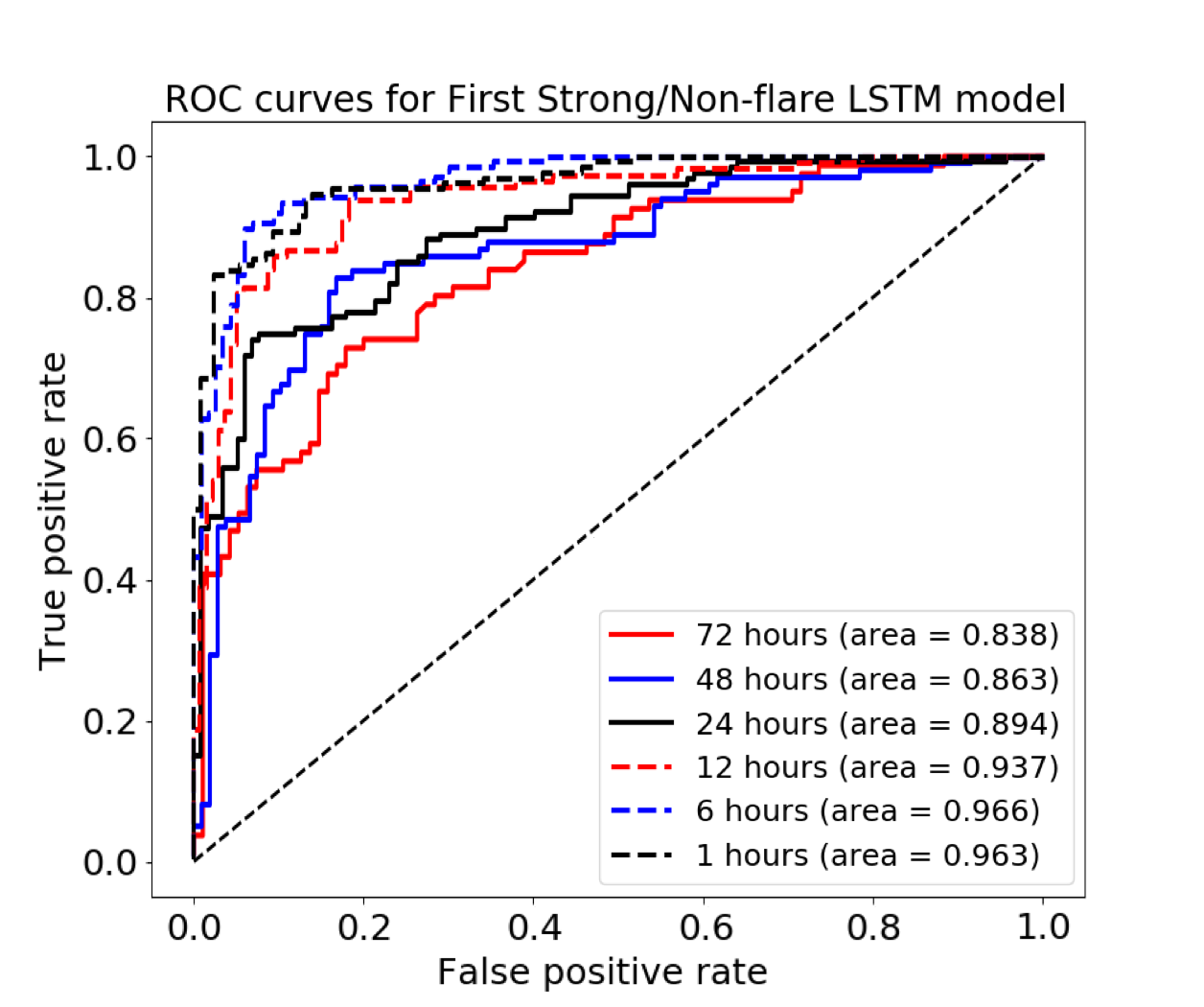}
\caption{ROC curve of LSTM model on M/X flare/non-flare classification with 1/3/6/12/24-hour prediction (left panel) and first M/x flare/non-flare classification with 1/6/12/24/48/72-hour prediction (right panel).}
\label{fig:fn_ROC}
\end{figure}

We use a two layer stacked LSTM architecture with $50$ cells in each layer. We choose a 50\% drop out rate in both layers to prevent over-fitting. The first LSTM layer provides a sequence output rather than a single output to feed into the second LSTM layer. A dense layer is added at the end with the sigmoid activation function that could generate a continuous value between 0 and 1 representing solar flare event probability. We utilize the binary cross-entropy as the loss function and the Adam optimization algorithm~\citep{kingma2014adam}. {We note that only in this subsection, flare/non-flare classification with SHARP parameters, we use 1 hour data for the LSTM models, which is a degenerate case since the input is a `time series' of one time point instead of multiple time points (used in later subsections).}
\tablename~\ref{table:first_flare_pred} gives the results for classifying ``solar flare event (of B/C/M/X class)'' against ``no solar flare event'' {1/3/6/12/24/48} hours prior to the start time of a solar flare event. See the left panel in \figurename~\ref{fig:fn_ROC} for corresponding ROC curves with AUC. 

\tabulinesep 0.5ex
\begin{table}[t!]
\centering
    \caption {First flare {(of any class)} classification results with $20$ SHARP parameters.}
\begin{tabu}{|c|ccccc|}
    \hline
    \rowcolor{lightgray} Metric & \multicolumn{5}{c|}{Number of hours before the first B/C/M/X flare} \\ 
    \rowcolor{lightgray}  & 1h & 3h & 6h & 12h & 24h \\
    \hline
    Precision  & 0.72 & 0.73 & 0.71 & 0.69 & 0.68 \\
    Recall  & 0.69 & 0.71 & 0.68 & 0.66 & 0.48 \\
    $F_1$ Score  & 0.70 & 0.72 & 0.69 & 0.67 & 0.55 \\
    ${\rm HSS}_1$ & 0.41 & 0.45  & 0.39 & 0.36 & 0.24 \\
    ${\rm HSS}_2$ & 0.43 & 0.45 & 0.39 & 0.36 & 0.25 \\
    TSS  & 0.43 & 0.45 & 0.40 & 0.36 & 0.25\\
    \hline
    \end{tabu}
    \label{table:first_flare_pred}
\end{table}


We also train an LSTM model that predicts strong flares (M/X class) from quiet times, which are hard to distinguish from B flares. The positive class is sampled from exactly {1/3/6/12/24/48/72} hours before the first strong flare event, and the negative class is sampled randomly from the time period of 48 hours prior to the first M/X flare event. \tablename~\ref{table:first_strong_flare_pred} gives the detailed results, {where metrics, such as precision, are higher} than those in \tablename~\ref{table:first_flare_pred}, which makes intuitive sense because it is much easier to tell strong flares from quiet times rather than weak flares from quiet times.

\tabulinesep 0.5ex
\begin{table}[htb]
\centering
    \caption {First strong flare {(M/X class)} classification results with $20$ SHARP parameters.}
\begin{tabu}{|c|ccccccc|}
    \hline
     \rowcolor{lightgray} Metric & \multicolumn{7}{c|}{Number of hours before the first strong flare} \\ 
    \rowcolor{lightgray}  & 1h & 3h & 6h & 12h & 24h & 48h & 72h \\\hline
    Precision  & 0.93 & 0.93 & 0.91 & 0.92 & 0.89 & 0.88 & 0.86\\
    Recall  & 0.88 & 0.87 & 0.85 & 0.85 & 0.77 & 0.72 & 0.68\\
    $F_1$ Score  & 0.90 & 0.90 & 0.88 & 0.88 & 0.83 & 0.79 & 0.76\\
    ${\rm HSS}_1$ & 0.81 & 0.80  & 0.77 & 0.77 & 0.68 & 0.62 & 0.57\\
    ${\rm HSS}_2$ & 0.81 & 0.79 & 0.77 & 0.77 & 0.68 & 0.62 & 0.56\\
    TSS  & 0.81 & 0.80 & 0.77 & 0.77 & 0.68 & 0.62 & 0.56\\
    \hline
    \end{tabu}
    \label{table:first_strong_flare_pred}
\end{table}


As we can see in \figurename~\ref{fig:fn_ROC}, the closer to the event time, the better the classification. Moreover, the event is much more predictive within 12 hours before the event. The rapid rise in predictive performance is consistent with the evolutionary timescale of ARs and suggests that within a period of $12-24$ hours, there is an observational signature indicating that a physical threshold has been passed at which point the flare becomes inevitable.  An example of such behavior is suggested by \citet{Schrijver:2007} who noted M and X flares occurring within 24 hours for ARs that have attained $10^{21}$ Mx of unsigned flux within 15 Mm of a strong polarity inversion line.  This further suggests that physical processes lead to a catastrophic loss of equilibrium following a buildup of energy, as has been suggested for a number of CME models \cite[cf.][]{Forbes:1991, Manchester:2003}. For periods longer than 24 hours, from the available observations, it may be physically impossible to make flare predictive classifications with high accuracy.


Furthermore, we train an LSTM model to predict, 24 hours ahead of time, whether an M/X flare occurs as opposed to no flare, as in~\citet{bobra2015solar}. The data are processed similarly as in~\citet{bobra2015solar}. All data are sampled from the $208$ ARs that produced M/X solar flare events. The positive class is sampled exactly $24$ hours prior to the time of the peak intensity of the event, and the negative class is sampled randomly from the period that no flare event would happen in the next 1/3/6/12/24/48 hours. 
\tablename~\ref{table:strong_no_flare} gives the detailed results. As we can see from \tablename~\ref{table:strong_no_flare}, the farther away from the M/X class event the negative class is selected, the better classifications we can get: the farther away from the M/X event, the ``quieter'' the region is in the negative class, thus the discrepancy between positive and negative events is larger. The key difference between the results in \tablename~\ref{table:strong_no_flare} and \tablename~\ref{table:first_strong_flare_pred} is how the negative class is determined/sampled, though both of them are aimed at predicting strong flares from non-flares. The sample selection mechanism behind \tablename~\ref{table:first_strong_flare_pred} shall give worse classifications but is less restrictive for the negative class as compared to the sample selection mechanism behind \tablename~\ref{table:strong_no_flare}. {These results again confirm our earlier comment that sample selection mechanism is important and it is essential to detail it for reproducibility of ML results.}

\tabulinesep 0.5ex
\begin{table}[htb]
\centering
    \caption{\linespread{0.75}\selectfont{}
    Strong Flare/Non-Flare $24$-hour ahead of event classification results with $20$ SHARP parameters. Each column represents the different mechanisms of selecting the negative class: no flare event happens in 1/3/6/12/24/48 hours.}
\begin{tabu}{|c|cccccc|}
    \hline
    \rowcolor{lightgray} 
     Metric & \multicolumn{6}{c|}{Selection Mechanisms of the Negative Class} \\ 
    \rowcolor{lightgray} 
     & 1h & 3h & 6h & 12h & 24h & 48h \\\hline
    Precision  & 0.89 & 0.90 & 0.90 & 0.90 & 0.93 & 0.95 \\
    Recall  & 0.79 & 0.79 & 0.80 & 0.82 & 0.87 & 0.90 \\
    $F_1$ Score  & 0.84 & 0.84 & 0.84 & 0.86 & 0.90 & 0.93 \\
    ${\rm HSS}_1$ & 0.69 & 0.70  & 0.71 & 0.73 & 0.81 & 0.86 \\
    ${\rm HSS}_2$ & 0.69 & 0.70 & 0.71 & 0.73 & 0.80 & 0.86 \\
    TSS  & 0.69 & 0.70 & 0.71 & 0.73 & 0.80 & 0.86 \\
    \hline
    \end{tabu}
    \label{table:strong_no_flare}
\end{table}


\subsection{Strong/Weak Flare Classification {with SHARP Parameters}}
\label{subsec:strong_weak_classification}

The Flare/Non-Flare model trained in \sectionname~\ref{subsec:flare_no_flare} predicts whether a flare is happening or not. Next, we train a model that classifies whether it is a strong flare (M/X class) or a weak flare (B class), given that a flare is happening. Note that we exclude the C flares here due to the fact that C flares could be arbitrarily close to strong B flares or weak M flares, making the classes highly indistinguishable. We first show the results of classifying M/X flares versus B flares using the SHARP parameters, and then the results using features obtained via the autoencoder followed by feature selection  See \sectionname~\ref{subsec:machine_learning_intro} for detailed descriptions of the algorithms. 

{In total, as recorded in the GOES data set, we have 751 strong flares and 3842 weak flares (see \tablename~\ref{tab:strong_weak_flares_years}).}
As mentioned in \sectionname~\ref{sec:method}, there are multiple flare events per AR and the flare events sometimes can be close to each other in time. To make sure that the {time series of the} flares are not overlapping in the training data, so that we are not using the same data point twice, we need to further prepare the data for training and testing by eliminating the overlapping events (see~\sectionname~\ref{subsec:pos_neg_prepare}). {The principle that we follow is to keep as many strong flares (the rarer class) as possible and randomly select one when two flares of the same class ``overlap''.} {Finally, see \tablename~\ref{table:no_events_strong_weak} for the detailed numbers of flare events and ARs corresponding to different number of hours before the first strong flare and the number of hours of data used to train and test the model.}


\tabulinesep 0.5ex
\begin{table}[htb]
\centering
    \caption{Number of flare events and ARs corresponding to different number of hours before the first strong flare and the number of hours of data used to train and test the model.}
\footnotesize
\begin{tabu}{|c|cccc|cccc|cccc|}
    \hline
 \rowcolor{lightgray} Hours Before an Event &	\multicolumn{4}{c|}{1 hour} &\multicolumn{4}{c|}{6 hours} & \multicolumn{4}{c|}{12 hours} \\
Hours of Data for Training &	1&		6	&	12	&	24	&	1	&	6	&	12&		24&		1	&	6	&	12	&	24\\
Num. Strong Flares & 585 & 579 & 565 & 543 & 579 & 565 & 559 & 529 & 565 & 559 & 546 & 510 \\
Num. Weak Flares & 851 &838 & 814 &768 &838	&817 & 794 & 749 & 814 & 794 & 769 &726\\
Num. ARs & 632& 628 & 618 & 606 & 628 & 619 & 612 & 601 & 618 & 612 & 608 & 588	\\
\hline
\rowcolor{lightgray} Hours Before an Event &\multicolumn{4}{c|}{24 hours} 	&\multicolumn{4}{c|}{48 hours} 	&\multicolumn{4}{c|}{72 hours} \\
 Hours of Data for Training &	1&		6	&	12	&	24	&	1	&	6	&	12&		24&		1	&	6	&	12	&	24\\
Num. Strong Flares &543&	529&	510	&480&475&463	&453	&423	&422	&412&	403&382\\
Num. Weak Flares &	768&	749&	726&	669&	660&	631&	609&	564&	560&	545&	524&	476\\
Num. ARs &606	&601	&588	&567	&563	&552	&542	&520	&518	&512	&504	&485\\
\hline
    \end{tabu}
    \label{table:no_events_strong_weak}
\end{table}

\tablename~\ref{table:strongweak} gives the strong and weak (M/X versus B class) flare classification results with $20$ SHARP parameters described in \sectionname~\ref{subsec:pipeline}. We use $12$ hours of data $t$ hours before an event, at a $1$ hour cadence, to classify the flare events; $t=1/6/12/24/48/72$ hours, corresponding to the last six columns in the table.

\tabulinesep 0.5ex
\begin{table}[htb]
\centering
    \caption{\linespread{0.75}\selectfont{} Strong and weak flare classification results from the LSTM model trained with 12 hours of data 1/6/12/24/48/72 hours (corresponding to the last six columns) prior to the flare event, using $20$ SHARP parameters. }
\begin{tabu}{|c|cccccc|}
\hline
\rowcolor{lightgray} Metric & \multicolumn{6}{c|}{Number of Hours before Event} \\ 
\rowcolor{lightgray} & 1h & 6h & 12h & 24h & 48h & 72h \\\hline
    Precision & 0.90 & 0.89 & 0.89 & 0.88 & 0.83 & 0.79 \\
    Recall & 0.86 & 0.84 & 0.81 & 0.77 & 0.73  & 0.76 \\
    $F_1$ Score  & 0.88 & 0.86 & 0.85 & 0.82 & 0.77 & 0.77 \\
    ${\rm HSS}_1$ & 0.76 & 0.73  & 0.70 & 0.67 & 0.57 & 0.56 \\
    ${\rm HSS}_2$ & 0.79 & 0.77 & 0.74 & 0.71 & 0.62 & 0.59 \\
    TSS  & 0.79 & 0.77 & 0.74 & 0.70 & 0.61 & 0.59 \\
    \hline
    \end{tabu}
 \label{table:strongweak}
\end{table}

\begin{figure}[tbph]
\centering
\includegraphics[width=1\textwidth]{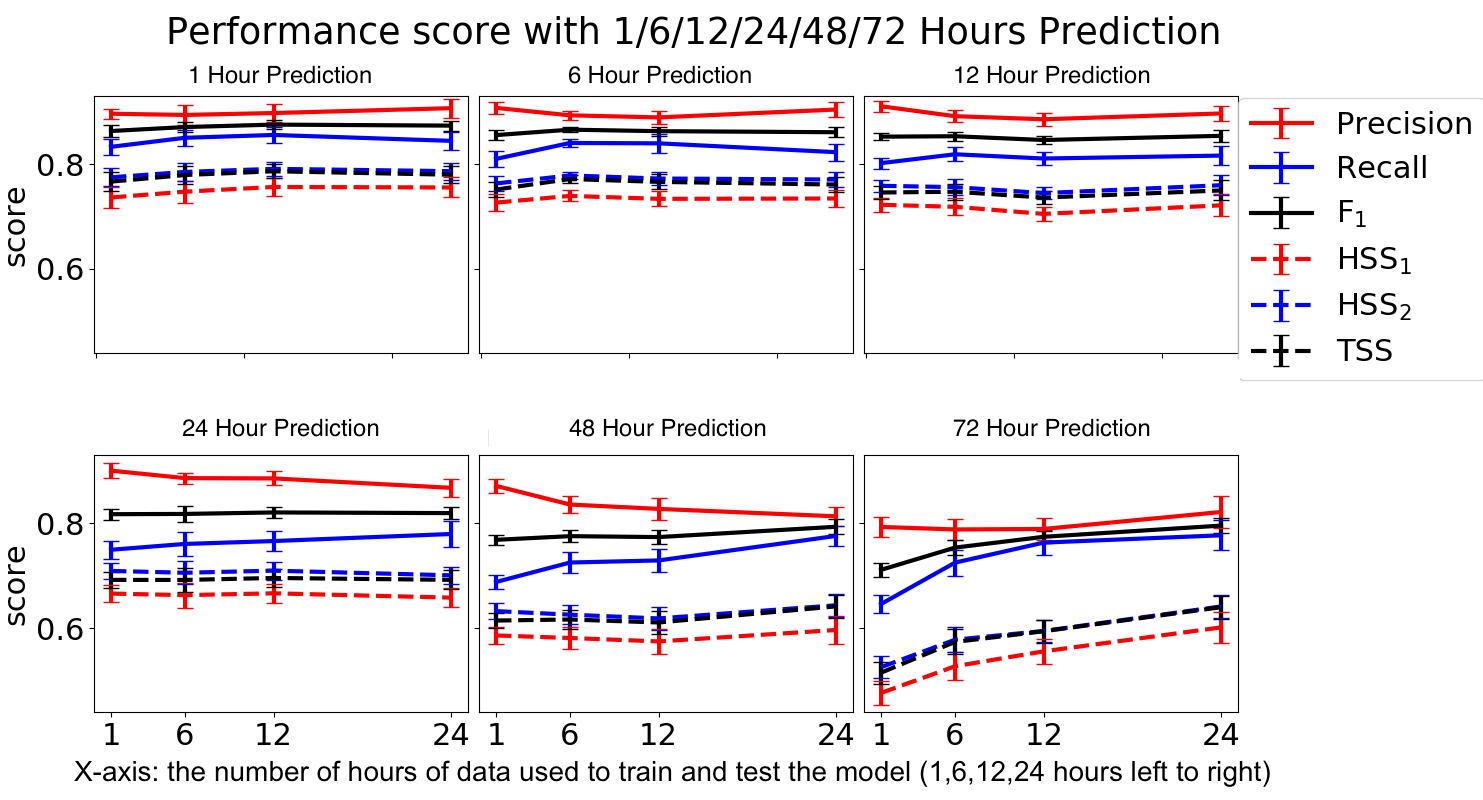}
\caption{The performance metrics on strong and weak flare event classification using LSTM with 20 SHARP parameters from HMI/SDO header file. {For each panel, the individual titles gives the forecasting window, i.e. number of hours' prediction. The x-axis for every panel, shared by the upper and lower panels, is the number of hours of data (1, 6, 12, 24 hours from left to right) used to train and test the model.}}
 \label{fig:sw_f1}
 \end{figure}
\begin{figure}[tbph]
\centering
\includegraphics[width=0.49\textwidth]{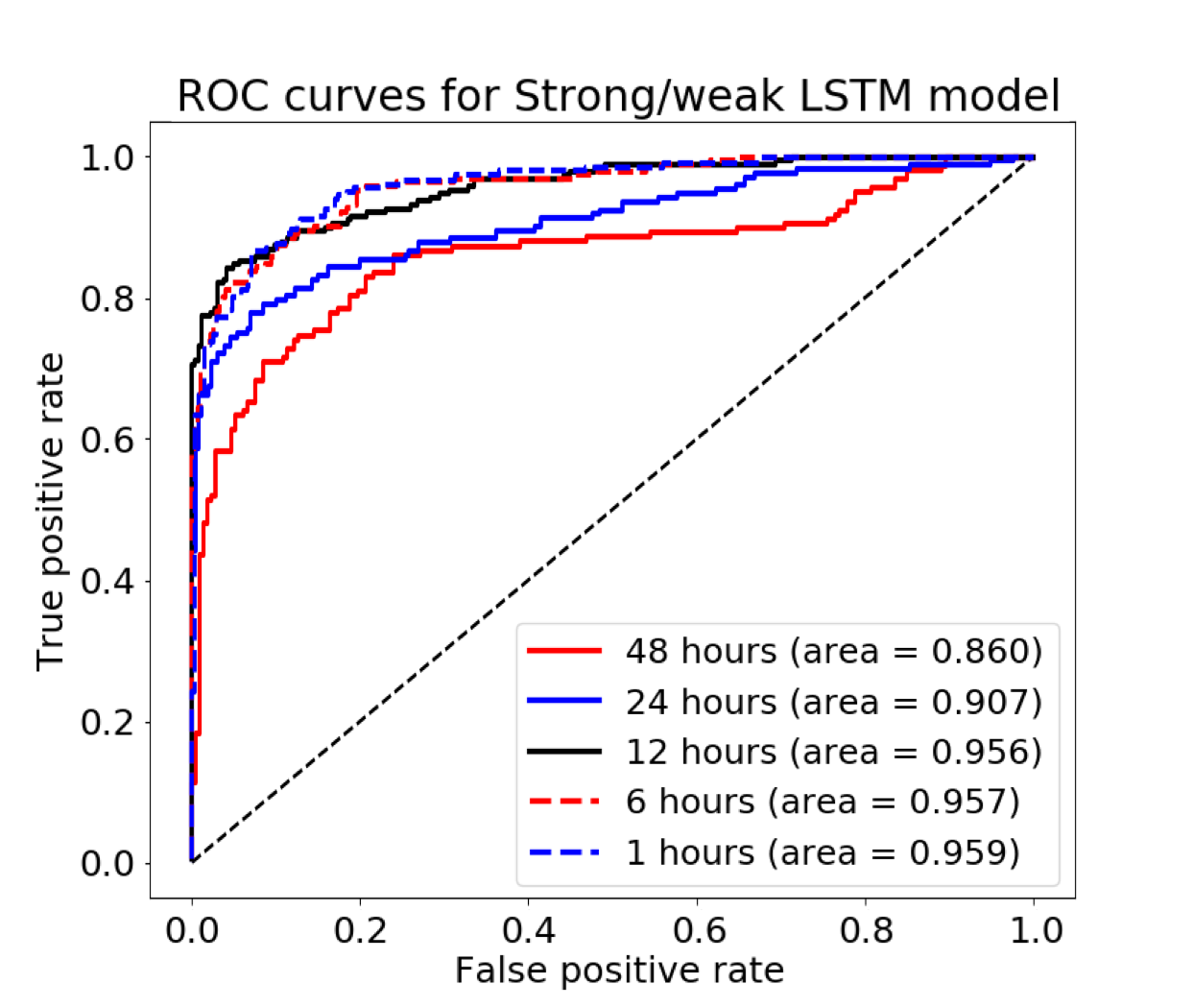}
\includegraphics[width=0.49\textwidth]{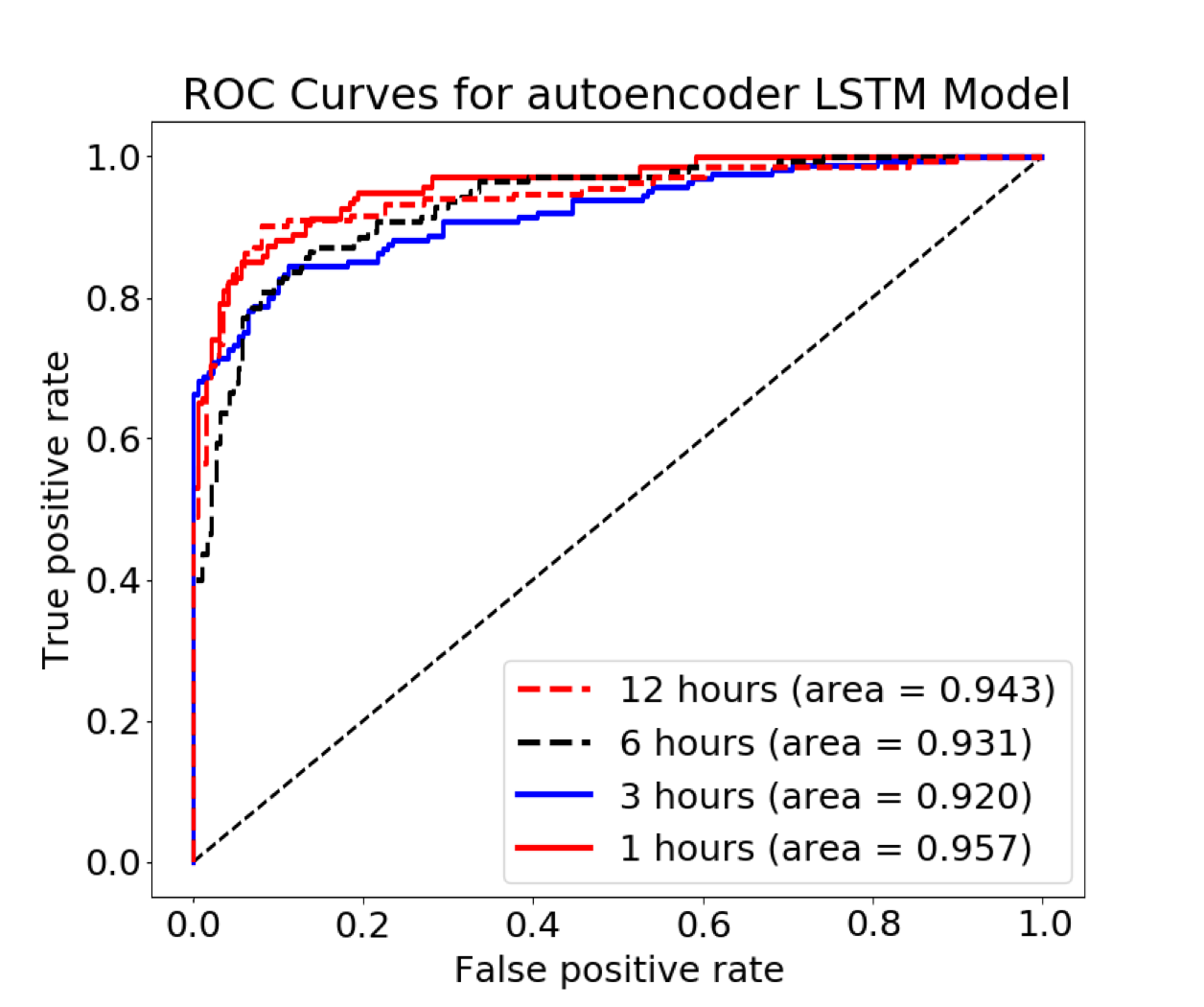}
\caption{ROC curve of LSTM model using 20 SHARP parameters (left panel) and machine-learned features using autoencoder (right panel) for strong/weak flare event classification (1/6/12/24/48 hours prior to event labeled with different colors and line types) with 1 hour data.}
\label{fig:sw_ROC}
\end{figure}

\figurename~\ref{fig:sw_f1} compares the $F_1$ score and other metrics for strong/weak flare classification. {We describe the rough trend that we observe based on the results given in \figurename~\ref{fig:sw_f1} while we acknowledge that these trends have not been verified rigorously due to the fact that different samples are used to train/test for different forecasting windows in this work.} Overall, the classification accuracy {appears to be} lower when predicting longer time ahead of an event. This is also exemplified in the ROC curves and AUC (area under the ROC curve) values given in the left panel of \figurename~\ref{fig:sw_ROC}, in which one hour's data is used for 1/6/12/24/48 hours' predictions. The AUC values of 48-hour prediction is much smaller than 24 hours' predictions, both of which are much smaller than 1/6/12 hours' predictions, where the latter three are not significantly different from each other. 

\subsection{{Feature Importance for Strong/Weak Flare Classification}}
\label{subsec:variable_importance}

Next we examine how these $20$ SHARP parameters contribute to the classification model. This is related to the notion of \textit{variable importance}, which is a widely adopted measure that represents the statistical significance of each feature in a model~\citep{garson1991interpreting,goh1995back}. Recall from \sectionname~\ref{subsec:pipeline} that the SHARP parameters are not independent features: USFLUX, TOTUSJZ, TOTUSJH, TOTPOT are highly correlated (with correlations ranging from 0.87 to 0.99); MEANPOT, SHRGT45, MEANSHR, MEANGAM are highly correlated (with correlations ranging from 0.8 to 0.99); SAVNCPP and ABSNJZH are highly correlated (with correlation 0.95); MEANALP and MEANJZH are highly correlated (with correlation 0.96); MEANGBZ and MEANGBT are highly correlated (with correlation 0.99). For these highly correlated features, as long as one of them is picked up as ``important'', all of the highly correlated ones are almost equally ``important''. Note that in the situation with highly correlated features, variable importance could become highly unstable. We take the backward elimination method as an example. In {each training/testing cycle of} backward elimination, we begin with all the features and delete one feature at each step, till all features are eliminated. Which feature is being deleted at each step can be determined by an exhaustive search of which one, among the remaining ones, upon removal, incurs the largest performance drop. However, when features are highly correlated, the resulting selected ``important'' features are not stable {across different training/testing cycles}: for two highly correlated features, one of them might be identified as ``important'' and the other identified as ``unimportant'' by the backward elimination method. 

To address the feature importance problem and  mitigate the difficulties incurred by the high correlations, we divide the 20 features into four groups, where features within each group are highly correlated with each other. {The dividing of the groups is based on the block structure in the correlation matrix of the 20 features, as shown in \figurename~\ref{fig:corr_parameters}, which have some physical similarities.} {Group 1 contains USFLUX, TOTUSJZ, TOTUSJH, TOTPOT and USFLUX, which are the total unsigned magnetic flux, electric current and current helicity and total potential energy, respectively. The latter three quantities are representative to differing degrees of the magnetic free energy.  Group 2 contains SAVNCPP and ABSNJZH, which are the net electric current per polarity and the absolute value of the net current helicity. These quantities are distinguished as integrated absolute values of the current and current helicity.  Group 3 contains three similar measures of AR area: SIZE ACR, NPIX and SIZE, but also contains NACR (number of strong magnetic-field pixels in the patch), which is more representative of magnetic flux. Group 4 contains features representative of the average density of the free energy.} These four groups are determined based on diagonal blocks in the correlation table (see~\figurename~\ref{fig:corr_parameters}).

\begin{figure}[tbph]
    \centering
    \includegraphics[width=0.9\textwidth]{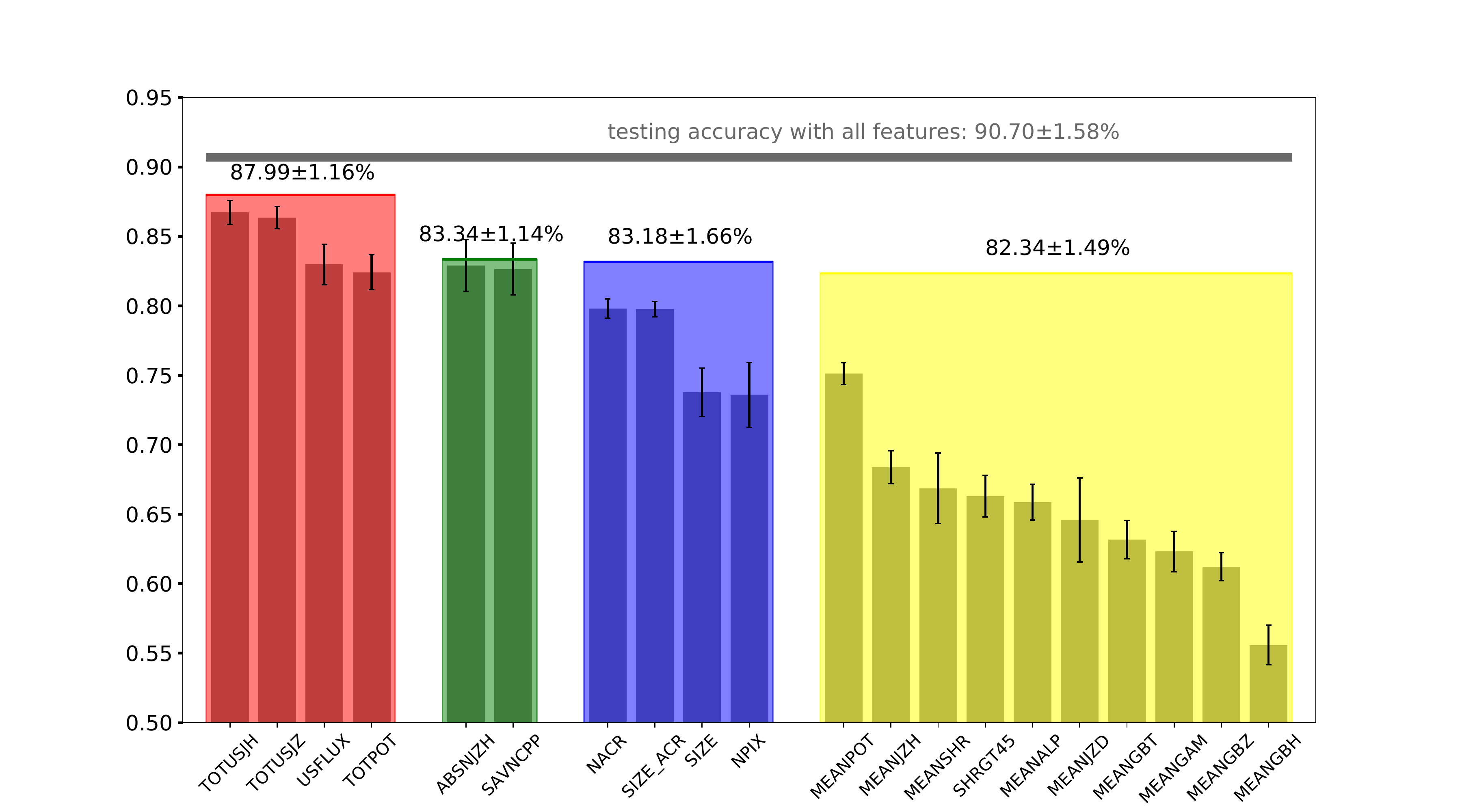}
    \caption{Feature importance considering correlations among features for the 6-hour ahead strong/weak flare classification using 24-hour long time series of SHARP features. The testing accuracy with all features is $90.70\pm 1.58\%$. The four groups of correlated features are labeled with red, green, blue and yellow colors, respectively, where on top of each colored block, the testing accuracy using the corresponding group of features alone is given. Each individual bar, together with the vertical black error bar, corresponds to the testing accuracy when we include only one feature in the LSTM model. }
    \label{fig:results_feature_importance}
\end{figure}

We explain our methodology via a concrete example, strong/weak flare classification using 24 hours' data (time series of SHARP parameters) for 6-hour predictions, as illustrated in \figurename~\ref{fig:results_feature_importance}. We begin with the LSTM with all of the features, which gives a baseline testing accuracy, 90.70\%, as shown by the gray horizontal line in \figurename~\ref{fig:results_feature_importance}. {Here the accuracy refers to the total number of correctly classified events divided by the total number of events (in the testing set).} We train the LSTM model with only one group of features at a time and report the corresponding accuracy for the four groups, which are {$87.99\pm 1.16\%$, $83.34\pm 1.14\%$, $83.18\pm 1.66\%$, and $82.34\pm 1.49\%$}, respectively; see the red, green, blue and yellow blocks in \figurename~\ref{fig:results_feature_importance}. Finally, we train the LSTM model with each feature alone, and report the corresponding testing accuracy, see the individual bars corresponding to each feature in \figurename~\ref{fig:results_feature_importance} and their error bars given by the black vertical bars, obtained through training the model with each feature 20 times with different random seeds.

We can see from \figurename~\ref{fig:results_feature_importance} that TOTUSJH (total unsigned current helicity, which indicates that the energy buildup due to the twist and shear of the magnetic field provides the energy erupted by the flares) and SAVNCPP (sum of the modulus of the net current per polarity) are important features for constructing precursors for strong solar flare events, {which confirms earlier studies}. Of course, the features that are highly correlated with these two features can be considered as ``almost equally important''. This result is consistent with alternative methods that we tried on variable importance quantification, {including the backward elimination~\citep{gregorutti2017correlation} and simple hypothesis testing methods~\citep{saeys2007review}.} We do not detail these alternative procedures since they give the same conclusions as the one described above.

\subsection{{Strong/Weak Flare Classification with Machine-Derived Features}}
\label{subsec:strong_weak_machine_features}

\begin{figure}[tbph]
    \centering
    \includegraphics[width=0.7\textwidth]{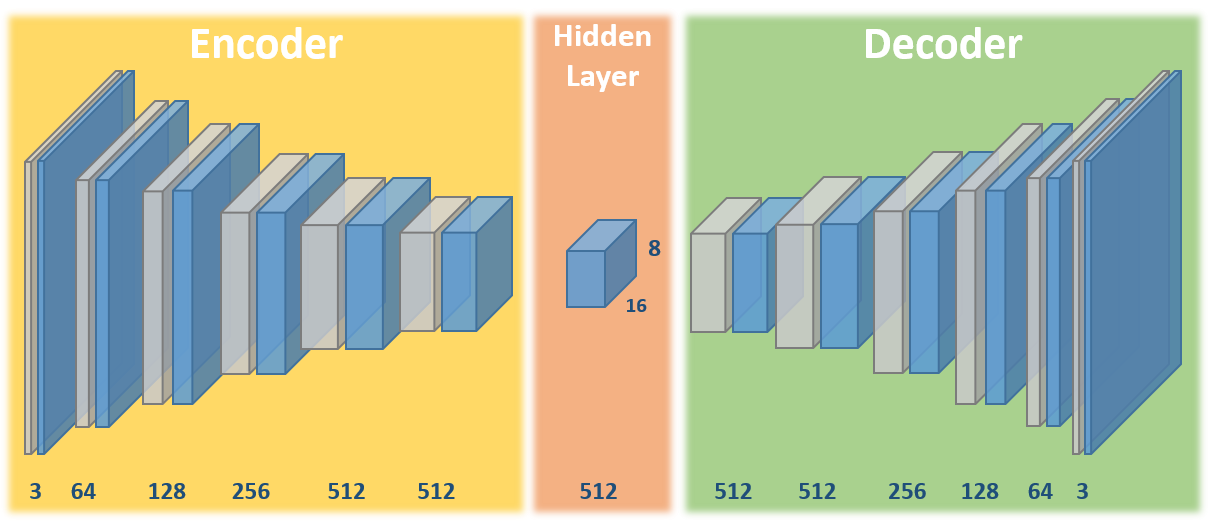}
    \caption{\linespread{0.75}\selectfont{} Structure of autoencoder on HMI images (3 components of the magnetic field). The numbers at the bottom corresponds to the dimensions at the encoding and decoding layers. We elaborate how we convert the HMI images to the final hidden layer (and reconstruct the HMI images using this hidden layer) of size $512\times 16\times 8$.}
    \label{fig:auto_encoder_hmi}
\end{figure}

In place of using the SHARP parameters, we will attempt to use the features extracted by a machine learning algorithm from the raw magnetic field images directly. Potentially this could give essential insight toward building new important features for solar flare predictions. We perform feature extraction via the autoencoder, as described in \sectionname~\ref{subsec:machine_learning_intro}. This is inspired by the VGG-16 architectures \citep{simonyan2014very} with a total of 20 layers (10 layers for encoder and 10 layers for decoder). The building blocks are:
\begin{enumerate}
\item 
a convolution layer (kernel size $3\times 3$, with same padding), the resulting output is of the same dimension with user specified number of channels,
\item
a max pooling layer (pooling size $2\times 2$ with stride $2\times 2$, and same padding), the resulting output is of half the dimension with the same number of channels, and 
\item
an unpooling layer (resizing image through bilinear interpolation), the resulting output is of user specified dimension with the same number of channels. 
\end{enumerate}
The final pooling layer of the encoder resizes the encoded image linearly to a constant size $8\times 16\times 512$. Consequently, $65,536$ features are extracted from the input image, regardless of the input dimension of the image. This creates the same number of features for input images of any size, which makes subsequent machine learning algorithms much easier to implement. 
\figurename~\ref{fig:auto_encoder_hmi} illustrates the structure of the adopted autoencoder.

%
%

\begin{figure}[tbph]
    \centering
    \includegraphics[width=1\textwidth]{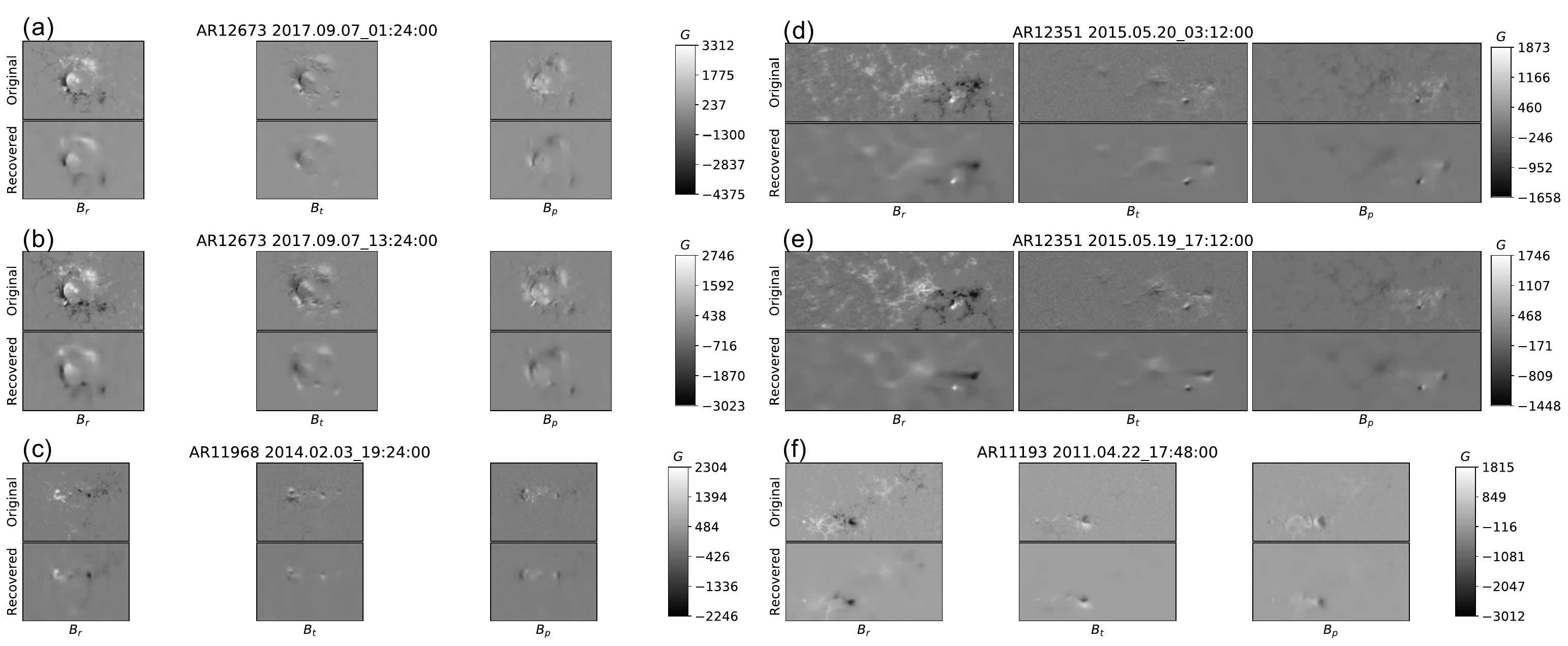}
\\ 
    \caption{{Demonstration of reconstructed images against original images (three components of the magnetic field data from HMI/SDO ARs, corresponding to the three columns in each panel) of several randomly selected ARs using the autoencoder. The AR numbers, dates (year.month.day), and times (hour:minute:second) of the images are given in the individual title of each panel. And the color scale on the right-hand-side of each panel reflects the strength of the three magnetic field components $B_r, B_t, B_p$ (in Gauss).} \vspace{1em}}
\vspace*{1em}
    \label{fig:autoencoder_reconstruction}
\end{figure}

Each input image is normalized before any encoding with the default Tensorflow image normalization, which effectively converts the data to mean 0 and standard deviation 1. Batch normalization \citep{ioffe2015batch} is applied for all the weights involved in convolution operations. For the activation function, we use the standard ReLu nonlinearity after each convolutional layer except for the final output layer. We add an additional $L_2$ regularization for all the convolution operations with tensorflow built in tuning for the hyperperameter $\lambda$. The initialization of weights are given by Gaussian random variables with mean 0 and standard deviation $10^{-3}$. This is a sensitive part of the algorithm that requires tuning. We adopt the Stochastic Gradient Descent (SGD) algorithm, the Adam Optimizer \citep{kingma2014adam}, with default coefficients, $\beta_1=0.9, \beta_2=0.999, \epsilon=10^{-8}$, where $\beta_1$ is the exponential decay rate for first moment estimate, $\beta_2$ is the exponential decay rate for the second moment estimate, and $\epsilon$ is a parameter for numerical stability. For the learning rate we initialize it to $0.01$, and decay it exponentially (by the scale of half) every 40 epochs. The loss function is given by Pixel by Pixel square difference across all channels: $\sum_{i,j,k} (x_{ij}^{(k)}-\hat{x}_{ij}^{(k)})^2$, where $x_{ij}^{(k)}$ is the pixel value of $k^{\rm th}$ channel at pixel index $i,j$, and $\hat{x}_{ij}$ is the reconstructed image. \figurename~\ref{fig:autoencoder_reconstruction} demonstrates the reconstructed images against the observed images of the three components of the magnetic field from HMI/SDO data, using several randomly chosen ARs. 

\begin{figure}[tbph]
\centering
\includegraphics[width=1\textwidth]{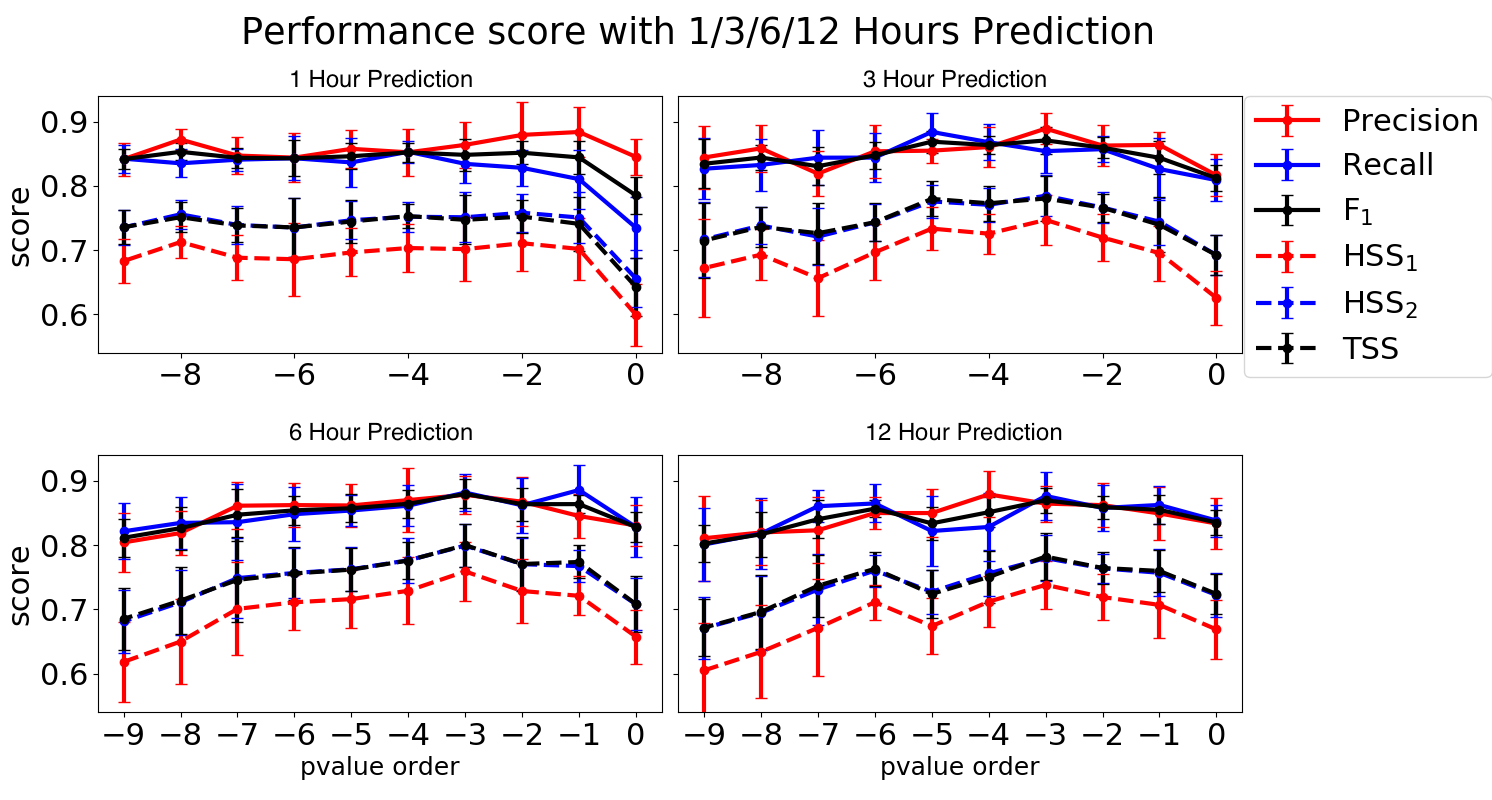}
\caption{Selection of threshold for $p$-values for marginal screening of features derived from autoencoders. For each panel, the x-axis is on the $\log_{10}$ scale of thresholds for $p$-values of selected features and the y-axis shows the corresponding metrics. {The corresponding number of features for the $p$-value orders from $-9$ to $0$ are $855, 1045, 1320, 1728, 2453, 3669, 5835, 10160, 20047, 65536$.}}
\label{fig:autoencoder_f1}
\end{figure}

As described in \sectionname~\ref{subsec:machine_learning_intro}, we need to perform feature selection prior to fitting the LSTM predictive classification model. The feature selection is based on marginally performing two-sample $t$-tests, and the thresholding $p$-value is a tuning parameter {based on cross-validation of performance scores that we choose}. {\figurename~\ref{fig:autoencoder_f1} shows the classification results using features selected from the autoencoder, with various thresholding $p$-values, corresponding to each forecasting window (number of hours ahead of events). We can see that the performance improves significantly with the feature selection as opposed to using all of the features from the autoencoder, which corresponds to the $p$-value threshold equal to $0$, the last column of each panel in \figurename~\ref{fig:autoencoder_f1}. For example, for $3$ hour prediction, we choose TSS as the performance score, which corresponds to the dashed black lines; then the $p-$value threshold $10^{-3}$, corresponding to $5,835$ features, gives the maximum TSS value. Therefore, we are able to reduce the number of features from $65,536$ to $5,835$ (more than 10 folds) with a much higher TSS score.}


Now we briefly explain why the performance for binary classification is improved after using the marginal screening method (based on $p$-values) to select a smaller number of features from all the 65,536 features given by the autoencoder. The p-values here are serving the purpose of ``identifying the useful features for strong/weak flare classification'' from the feature pool extracted from the autoencoder, which is actually deriving features to reconstruct the image. A significant p-value (the significance level is a tuning parameter) {indicates} the ``usefulness'' of the corresponding feature. In statistics, many redundant useless features could result in poor classification results, especially in the case that we are faced with: the number of features is much larger than the number of events (M/X or B flares) that we consider (see \sectionname~\ref{subsec:machine_learning_intro} for references). Therefore, this feature selection technique that we are using conveys two messages: first, we do not need so many features to achieve good performance; second, removing useless features actually improves the performance and suggests the possibility of identifying machine-derived physically meaningful features.


The right panel in \figurename~\ref{fig:sw_ROC} {in \sectionname~\ref{subsec:strong_weak_classification}} shows the ROC curve of strong/weak flare classifications using features derived from the autoencoder with feature selection $p$-value threshold set at $10^{-3}$. Different line types/colors correspond to 1/3/6/12 hours of prediction. Note that we only train the autoencoder with time series of 12 hours (data from 0-12 hours prior to an event with cadence 1 hour is used to train the autoencoder), thus we cannot make predictions longer than 12 hours. However, the LSTM model with the machine derived features can be readily adapted to any desired number of hours of forecasting window, similar to the LSTM models with SHARP parameters trained in Section~\ref{subsec:strong_weak_classification}. As we can see from \figurename~\ref{fig:sw_ROC}, the AUC for {1/6/12} hour predictive classifications are (0.959, 0.957, 0.956) with SHARP parameters and (0.957, 0.931, 0.943) with features derived from autoencoder.  This shows that the latter performs the same as if not worse than the former, according to AUC. Note that in the autoencoder model, the AUC is not monotonic as a function of the forecasting window since the marginal screening step, which is performed separately for each forecasting window, incurs extra heterogeneity.

\subsection{Case Study on Flare Classification}
\label{subsec:case_study}

\begin{figure}[htb]
\centering
\includegraphics[width=1\textwidth]{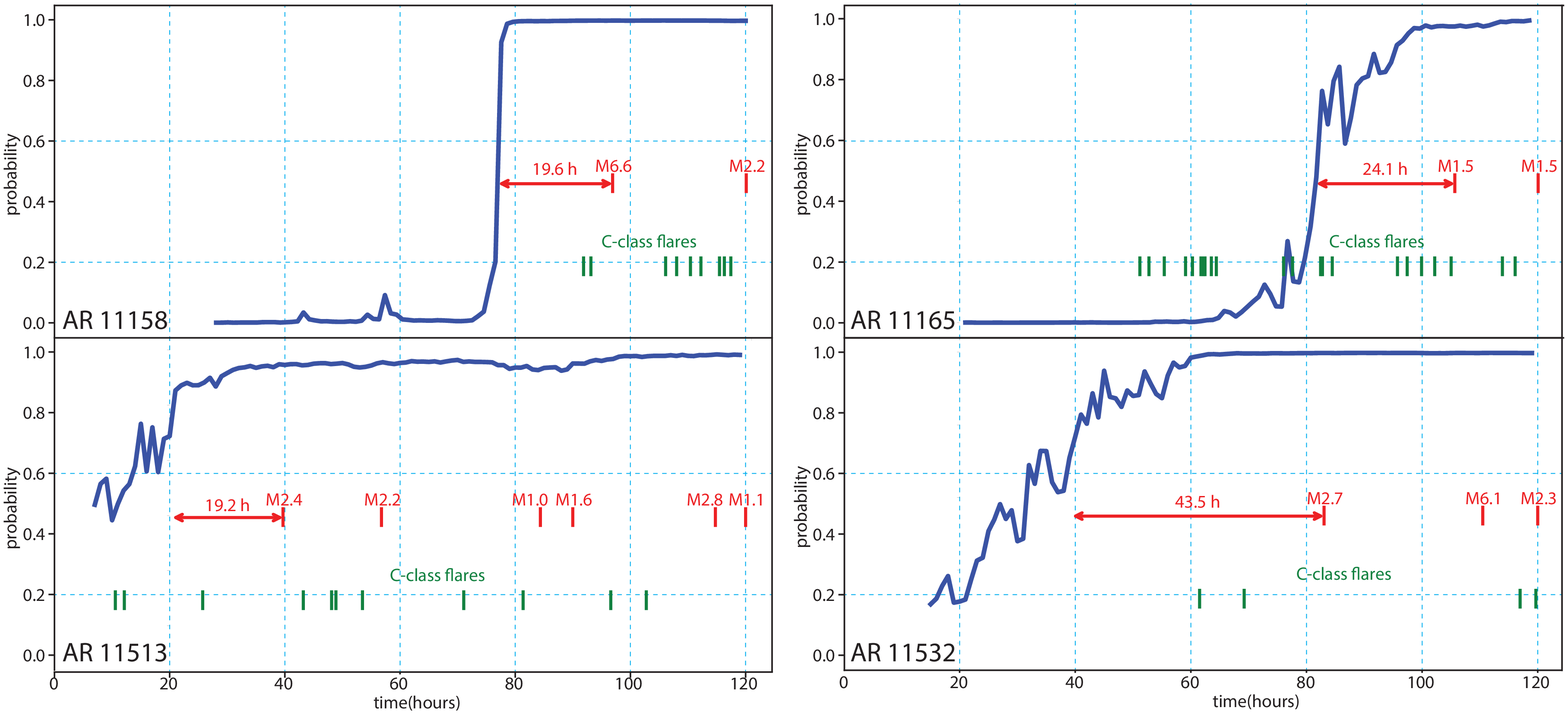}
\caption{Case studies on four ARs 120 hours prior to {the peak intensity time of M/X events at 2011-02-14 17:26:00 (AR11158), 2011-03-07 21:50:00 (AR11165), 2012-07-02 00:35:00 (AR11513), and 2012-07-29 06:22:00 (AR11532)}. Strong/Weak flare classification LSTM model is used to predict the probability (classification score) of a M/X class event happening at a specific time (blue curve) with observed C and M flare events with green and red colors, respectively. The classification scores go higher when we get closer to the M/X class event and {a sharp or gradual} transition of the classification score happens around a day ahead of the first strong flare.
}
\label{fig:case_study_1}
\end{figure}

We randomly choose four ARs (with NOAA AR numbers 11158, 11165, 11532, 11513) to show our LSTM model Strong/Weak flare  (\sectionname~\ref{subsec:strong_weak_classification}) classification scores time periods ranging from very beginning until the final strong, M/X class flare events (see \figurename~\ref{fig:case_study_1}). Note that in our data extraction pipeline, we do not fetch data from the period when strong and weak flare events heavily overlap (we do not consider this scenario yet in the current LSTM model). Thus the number of available ARs with long time range data before the M/X class event is not many. These classification scores, though obtained from a strong/weak flare classification model (instead of an operational flare prediction model), already show an increasing pattern as we approach around 20 hours prior to the final M/X class event.



Here are more details on model training and calculation of the classification scores. {Both the strong and weak flares are sampled $1$ hour prior to the flare event at a $1$ hour cadence, which gives $721$ strong flares and $721$ weak flares for training the LSTM model for strong/weak flare classification.} {Note that we use the same number of strong flares and weak flares (a simple random sample from all) here. This in fact gives a conservative demonstration of our algorithm: assuming no prior knowledge about the solar physics and no learned knowledge about the rareness of the strong events (i.e., the sample unbalance problem), we show how the ML algorithm we train can differentiate strong flares from others.} After training the LSTM models for strong/weak flare classification {(see Section~\ref{subsec:strong_weak_classification} for details of the structure of the LSTM model)}, we save the weight parameters and use them to predict scores (between $[0,1]$) representing the probability that there will be a (strong) flare event happening at each future time point by feeding the current data features into the {trained} model. These ``weight parameters'' actually refer to the trained nonlinear transformations of the SHARP features in the LSTM model. In essence, we save our trained model and use it as a black box for calculating the classification scores for the four ARs {that we test on}. 



In \figurename~\ref{fig:case_study_1}, we compare the sequence of classification scores (blue solid line) with the time of observed flare events (red for M flares and green for C flares) for each of the four ARs (with NOAA AR numbers 11158, 11165, 11513 and 11532) from the GOES data set to check the validity of the predictions, i.e. whether the classification scores increase prior to any (strong) flare event. {The end time of each case (AR) that we consider here is given by the peak intensity of M flares at 2011-02-14 17:26:00 (AR11158), 2011-03-07 21:50:00 (AR11165), 2012-07-02 00:35:00 (AR11513), and 2012-07-29 06:22:00 (AR11532).} 
We note that these four ARs were excluded from the training of the classification model. It should also be noted that due to the rotation of the sun, an AR cannot be seen for more than approximately 350 hours at a time. The 100 consecutive SDO/HMI features with a cadence of 1 hour cover a very significant fraction of this AR visibility.

\begin{figure}[htb]
    \centering
    \includegraphics[width=0.7\textwidth]{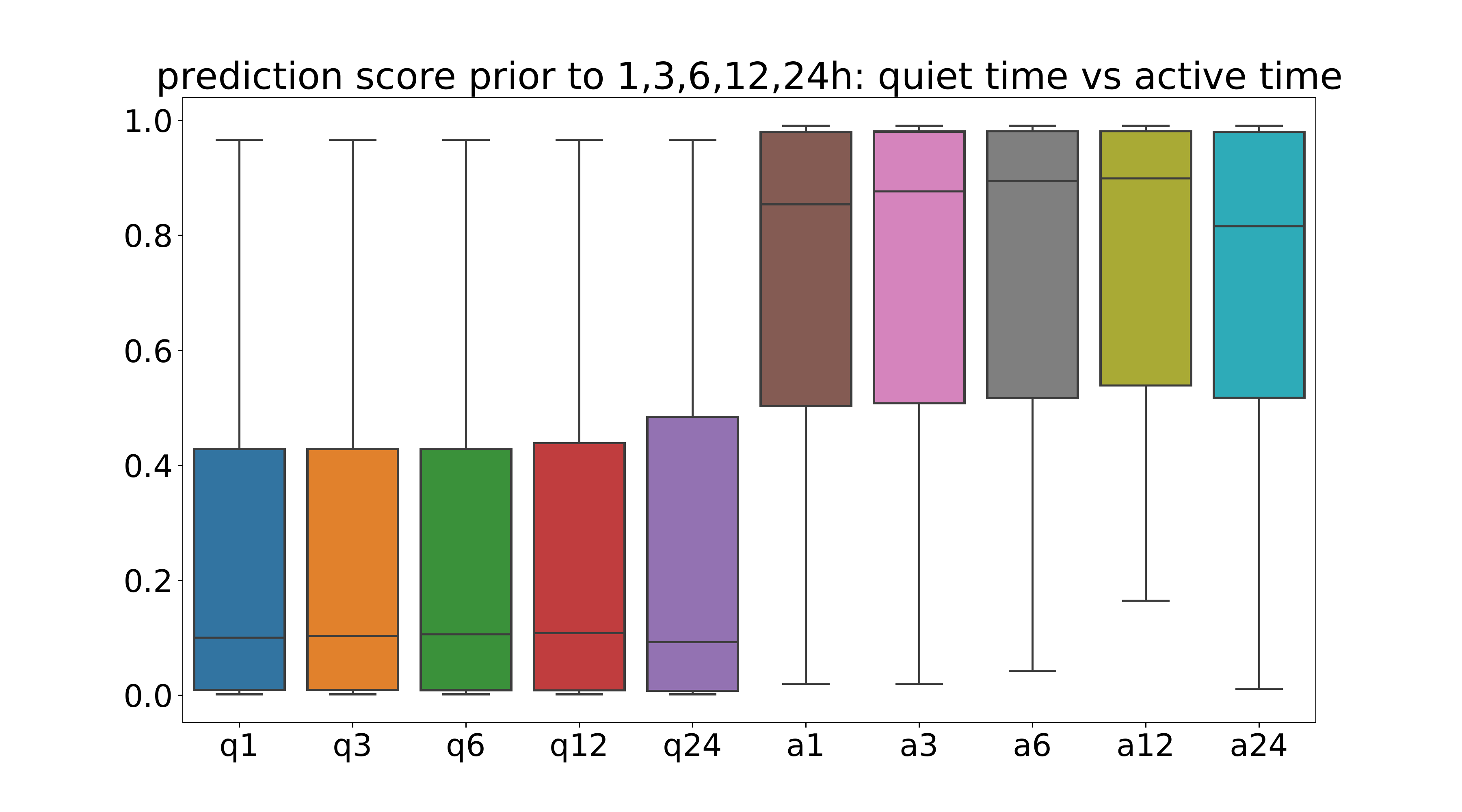}
    \caption{Boxplots of the classification scores for the case studies done for the four ARs over the entire observed time range. The X-axis label stands for q (quiet time, first five columns) or a (active time, last five columns) with [1,3,6,12,24] hours' predictions. The Y-axis label is the corresponding classification score. }
\label{fig:case_study_boxplot}
\vspace{2em}
\end{figure}

Furthermore, \figurename~\ref{fig:case_study_boxplot} shows box plots of the classification scores {1/3/6/12/24} hours prior to a ``quiet time'' (first five columns) and ``active time'' (time of peak intensity of strong flare events, last five columns), for the four ARs in the entire time range: year 2010 to year 2018. We define a certain time as ``quiet time'' if there is no strong flare before or after 24 hours. {We can see from this figure that the classification scores are well-separated by 0.5 for the ``quiet time'' and ``active time'', which further validates our construction of precursors for strong solar flare events using the LSTM model.}

Our preliminary results indicate that with the time-dependent learning process, the machine learning algorithm identified examples of a large gradient in the classification score approximately 20-24 hours before a large (M/X class) flare. At this point, we cannot translate this result to physical understanding of the flare initiation mechanism. This work will be the subject of a subsequent publication. {The result is highly encouraging in the sense that we seem to have shown the existence of some physical parameter combination that is capable of detecting strong flares by a significant time in advance for several ARs.} 

\section{Conclusions and Future Work}
\label{sec:conclusion}

{We have presented machine learning algorithms that give encouraging results in classification of strong and weak solar flare events and in detecting efficient precursors for strong flares, using the SDO/HMI vector magnetograms and/or SHARP parameters. This work serves as our first attempt toward early predictions of strong solar flare events.}

To summarize, we developed a flexible pre-processing pipeline to prepare data from multiple sources (GOES, HMI/SDO) for subsequent machine learning algorithms. Then we trained {the LSTM model} to perform two classification tasks: flare/no-flare and strong/weak flare classification. We use SHARP parameters primarily for the two classification models. Beyond using derived quantities, {i.e. SHARP parameters}, we apply the autoencoder to extract features directly from images of all components of the magnetic field. Feature selection is performed to get rid of redundant noisy features that may harm subsequent classifications. We then show that these machine-derived features can predict/classify almost as well as the SHARP parameters derived from physical understanding. 

Compared with previous results, our methodology and the results presented in this paper stand out in several aspects. 

\begin{enumerate}
    \item We train models with {1/3/6/12/24/48/72}-hour forecasting windows of flare events, instead of a single fixed forecasting window of 24 hours. We discover the interesting and physically meaningful phenomenon of the ``phase transition'' of around 24-hour predictions: for shorter forecasting windows, the performance of classification does not vary too much and for longer forecasting windows, the performance (or capability) of classification drops quite noticeably. This corresponds to the underlying physics: the energy build-up takes around 12 to 24 hours for a solar flare event, which we discuss in detail in \sectionname~\ref{subsec:flare_no_flare} (where the references are given). Further investigations will study the cause and effect of this ``phase transition phenomenon'', both from a physics perspective and a machine learning perspective. 

\item We train multiple models to perform a sequence of predictive classification tasks (M/X flare/weak flare classification), and finally combine them to obtain {encouraging} results. This has not been done before as far as the authors have been able to find in the literature. The decomposition of the challenging task of solar flare predictive classification into several smaller/easier tasks enabled us to assess the possibility and limitations of using HMI data for the precise classification of solar flare events. This serves as a great first step toward using more advanced machine learning and statistical analysis techniques to finally enable efficient and accurate real-time solar flare forecasting.

\item The modeling techniques that we use give us {high-quality} classification results in terms of HSS and TSS scores, metrics that are commonly adopted in the field. The LSTM model that we use for predicting the outcome of a time series observation not only takes care of the ``stationary features'' (which are the features adopted in {most of the} work in the literature, such as predictions using the SVM, random forest, penalized regression), but also takes care of the time evolution of features/images. 

\item {We use the autoencoders to automatically extract features from images, in addition to using physical quantities from the magnetograms. These quantities (SHARP parameters here) are derived from physical understanding and have been used successfully in many previous examples, e.g. \citet{Falconer:2001, Leka:2003a, barnes2007probabilistic, bobra2015solar}. It is very encouraging that our machine-derived features can be used to predict/classify almost as well as the SHARP parameters.  In fact, these parameters represent an incomplete understanding of solar flare events, which the autoencoder features may surpass.  First, the most valuable parameters for prediction in our study, SAVNCPP and TOTUSJH, are scalar values representing integrals of electric current and current helicity, respectively.  While much of the information regarding the spatial distribution of the magnetogram has been lost in these variables, it remains fully available to the autoencoded features. Refining the use of the autoencoder will be left for further investigations in our ongoing/future work.}


\item In our handful of case studies, the strong flare (M/X class) classification scores showed a sharp {(or gradual)} increase at least $20^{\rm h}-25^{\rm h}$ before the first large flare. This implies that there is a still unexplored (probably nonlinear) combination of the SHARP parameters that exhibits a runaway effect about a day before large solar flares. In the future we intend to further explore this exciting result from both the machine learning and physics perspective. It is our hope that eventually this discovery might lead to flare forecasts with lead times greater than one hour.
\end{enumerate}

{Our ongoing and future work includes (a) combining features from the Atmospheric Imaging Assembly (AIA) data with the current feature set, (b) connecting machine-learned features to derived quantifies (such as the SHARP parameters) to facilitate scientific discoveries of new physically meaningful features, and (c) training physically based machine learning models for accurate estimation of flare event time and flare event intensity. The last one will potentially lead to operational flare forecasting. }

\vspace{12pt}
 
\noindent
\textbf{Acknowledgements.} We thank Enrico Landi, Justin Kasper, Tuija Pulkkinen, Igor Sokolov and Bart van der Holst of the Department of Climate and Space Sciences and Engineering for helpful discussions. We also acknowledge the help of Monica Bobra (Stanford) and K.D. Leka (NWRA).  We also acknowledge the efforts of several UM master students recently involved in the project: Hu Sun, Zhenbang Jiao, Chung Hoon Hung, Boyang Zhang, and Bruce Park. This work was supported by NASA grants 80NSSC19K0373 and 80NSSC18K1208, NSF grant AGS-1322543, and by the Michigan Institute for Data Science (MIDAS) at the University of Michigan. All SHARP data used in this study are available from the Joint Science Operations Center (JSOC) NASA grant, see \url{http://jsoc.stanford.edu/}. All relevant digital values used in the manuscript (both data and model) will be permanently archived at the U-M Library Deep Blue data repository, which is specifically designed for U-M researchers to share their research data and to ensure its long-term viability. Data sets will be assigned Digital Object Identifiers (DOIs) which will serve as identifiers for the data, enabling them to be cited in publications.

\begin{appendices}





\section{Tables of Confusion Matrices}

We give confusion matrices~\citep{provost1998glossary}, i.e. list the numbers of TP (true positives), FN (false negatives), TN (true negatives) and FP (false positives), for the classification results in Sections~\ref{subsec:flare_no_flare} and~\ref{subsec:strong_weak_classification}. We run the machine learning algorithms 20 times with different seeds, thus the mean, minimum and maximum values are given in \tablename~\ref{table:confusion_flare_no_flare},~\ref{table:confusion_strong_noflare}, and~\ref{table:confusion_strong_weak}. This show the robustness and replicability of our results. 

\tabulinesep 0.5ex
\begin{table}[htb]
\centering
        \caption {\linespread{0.75}\selectfont{} Flare/Non-Flare classification confusion matrix with $20$ SHARP parameters. This corresponds to \tablename~\ref{table:first_flare_pred}.}
\begin{tabu}{|c|cccc|}
    \hline
    \rowcolor{lightgray} 
     Forecasting Window & \multicolumn{4}{c|}{Contingency Table (mean [min, max])} \\ 
    \rowcolor{lightgray} 
     & TP & FN & TN & FP  \\\hline 
1 hr & 53.0 [39,62] & 23.8 [12,34] & 60.0 [49,72] & 21.2 [15,33]\\
3 hr & 54.9 [49,66] & 22.6 [11,33] & 57.4 [51,64] & 20.2 [11,31]\\
6 hr & 51.1 [41,61] & 24.1 [17,33] & 53.5 [42,60] & 21.3 [11,33]\\
12 hr & 47.1 [40,54] & 24.3 [13,32] & 49.2 [40,57] & 21.5 [14,31]\\
24 hr & 29.4 [17,40] & 32.5 [16,50] & 47.8 [40,53] & 14.3 [5,25]\\
48 hr & 24.9 [15,34] & 16.1 [6,28] & 26.2 [19,34] & 13.9 [5,23]\\
    \hline
    \end{tabu}
        \label{table:confusion_flare_no_flare}
\end{table}


\tabulinesep 0.5ex
\begin{table}[htb]
\centering
        \caption {First Strong Flare/Non-Flare classification confusion matrix with $20$ SHARP parameters. This corresponds to \tablename~\ref{table:first_strong_flare_pred}.}
\begin{tabu}{|c|cccc|}
    \hline
    \hline
    \rowcolor{lightgray} 
     Forecasting Window & \multicolumn{4}{c|}{Contingency Table (mean [min, max])} \\ 
    \rowcolor{lightgray} 
     & TP & FN & TN & FP  \\\hline
    1 hr   & 113.3 [107,120] & 16.2 [11,24] & 120.9 [109,128] & 8.7 [3,16] \\
    3 hr   & 114.1 [102,125] & 17.8 [9,27] & 116.5 [106,127] & 8.7 [4,14] \\
    6 hr   & 106.7 [95,115] & 18.2 [13,24] & 117.6 [107,125] & 10.6 [5,18] \\
    12 hr  & 106.3 [91,118] & 19.0 [10,27] & 115.1 [100,125] & 9.7 [6,17] \\
    24 hr  & 93.1 [76,103] & 27.9 [20,39] & 112.2 [100,124] & 11.0 [5,19] \\
    48 hr  & 72.8 [63,79] & 28.2 [32,39] & 94.6 [83,106] & 10.4 [2,25]  \\
    72 hr  & 61.8 [54,74] & 28.9 [18,36] & 75.1 [68,82] & 10.2 [6,19]  \\
    \hline
    \end{tabu}
        \label{table:confusion_strong_noflare}
\end{table}

\tabulinesep 0.5ex
\begin{table}[htb]
\centering
        \caption {Strong/Weak flare classification confusion matrix with $20$ SHARP parameters. This corresponds to \tablename~\ref{table:strongweak}.}
\begin{tabu}{|c|cccc|}
    \hline
    \hline
    \rowcolor{lightgray} 
     Forecasting Window & \multicolumn{4}{c|}{Contingency Table (mean [min, max])} \\ 
    \rowcolor{lightgray} 
     & TP & FN & TN & FP  \\\hline
    1 hr   & 161.4 [144,176] & 27.3 [17,40] & 247.8 [230,265] & 18.6 [7,28] \\
    6 hr   & 153.4 [131,169] & 29.3 [22,45] & 244.1 [229,264] & 19.4 [12,28] \\
    12 hr  & 145.9 [133,161] & 34.1 [25,43] & 234.2 [216,250] & 18.9 [11,27] \\
    24 hr  & 128.5 [116,144] & 39.2 [27,57] & 221.6 [206,240] & 16.8 [9,27] \\
    48 hr  & 106.3 [90,118] & 39.5 [27,56] & 166.8 [155,191] & 22.5 [11,35]  \\
    72 hr  & 87.4 [80,101] & 27.2 [17,38] & 115.8 [99,123] & 23.7 [13,36]  \\
    \hline
    \end{tabu}
        \label{table:confusion_strong_weak}
\end{table}


\section{Additional Results}
\label{subsec:more_results}

In this Section, we give results of strong/weak flare classifications based on alternative sample-splitting methods described in Section~\ref{subsec:data_split}: split-by-active-region (including correcting for over-representation of certain highly flaring ARs) and split-by-year (that considers solar active phase and decaying phase). {For all the figures in this Section, ``prediction period'' refers to the number of hours prior to a flare event, i.e. $X$ hours prediction, with {$X=1/6/12/24/48/72$}. }

\begin{table}[ht!]
    \centering
        \caption{Proportion of positive class (strong flares) in training and testing data, in the format of mean $\pm$ standard deviation, for different cap values we specify in the split-by-active-region, see Section~\ref{subsec:data_split}.}
    \begin{tabu}{|c|cc|}
    \hline
\rowcolor{lightgray} Cap &	Training (Mean $\pm$ Std.) & Testing (Mean $\pm$ Std. )\\
\hline
2& 0.298 $\pm$ 0.019 & 0.669 $\pm$	0.139\\
3& 0.343 $\pm$ 0.023 & 0.741 $\pm$ 0.141\\
4& 0.399 $\pm$ 0.032 & 0.716 $\pm$ 0.174\\
5& 0.459 $\pm$ 0.029 & 0.608 $\pm$ 0.108\\
10& 0.569 $\pm$	0.044 & 0.651 $\pm$ 0.143\\
15&	0.619 $\pm$	0.043 &	0.673 $\pm$	0.113\\
$\infty$ & 0.693 $\pm$ 0.071 &	0.680 $\pm$ 0.146\\
\hline
    \end{tabu}
    \label{tab:train_test_prop_split_AR}
\end{table}

\enlargethispage{2ex}
\begin{figure}[ht!]
    \centering
    \includegraphics[width=0.9\textwidth]{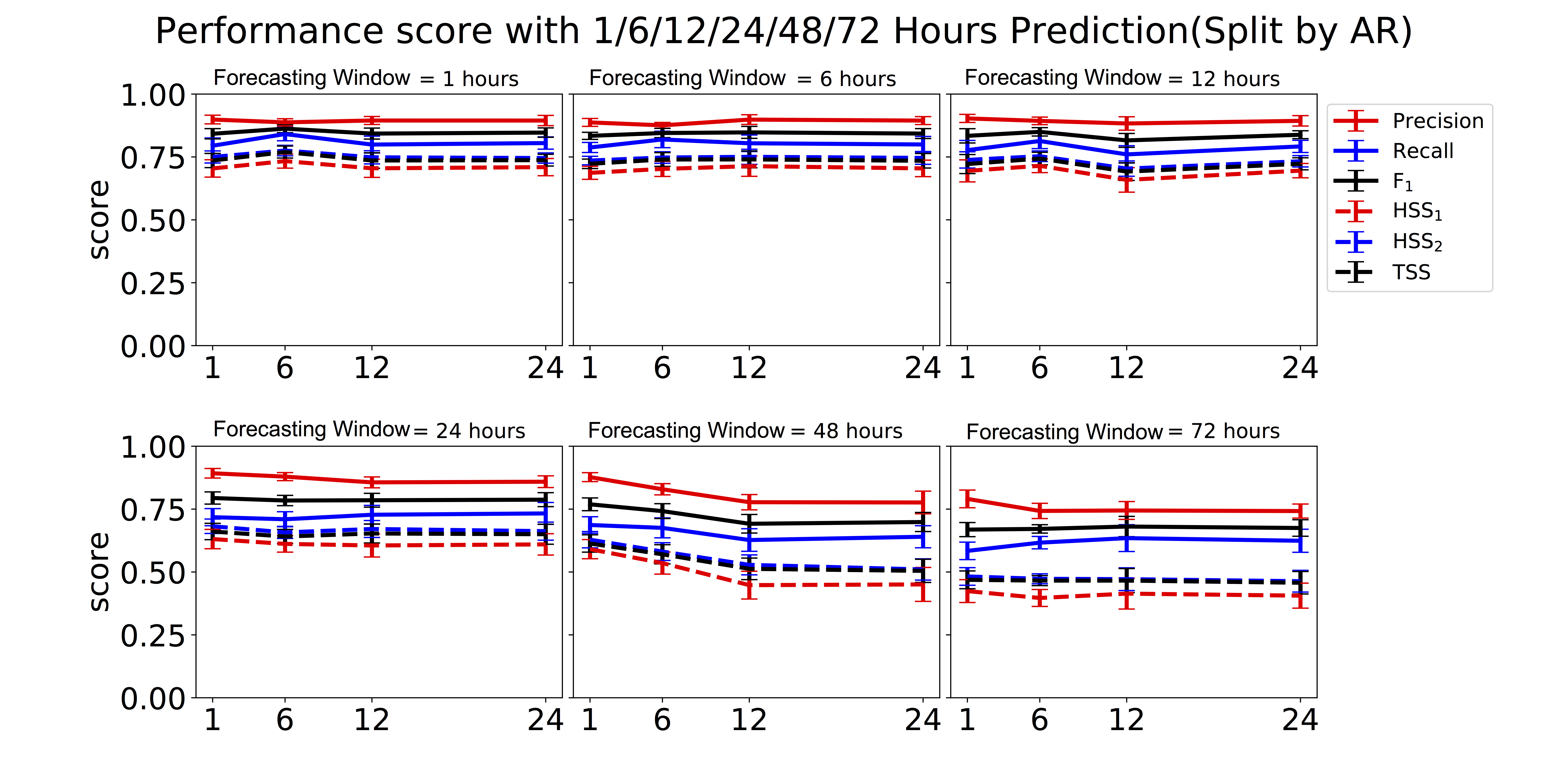}
    \caption{Performance scores from split-by-active-regions (with no cap on the number of events per AR), as described in Section~\ref{subsec:data_split}, are displayed in the same way as in \figurename~\ref{fig:sw_f1} in Section~\ref{subsec:strong_weak_classification} in the main text.}
    \label{fig:split_ar}
\end{figure}

\begin{figure}[htb]
    \centering
    \includegraphics[width=0.9\textwidth]{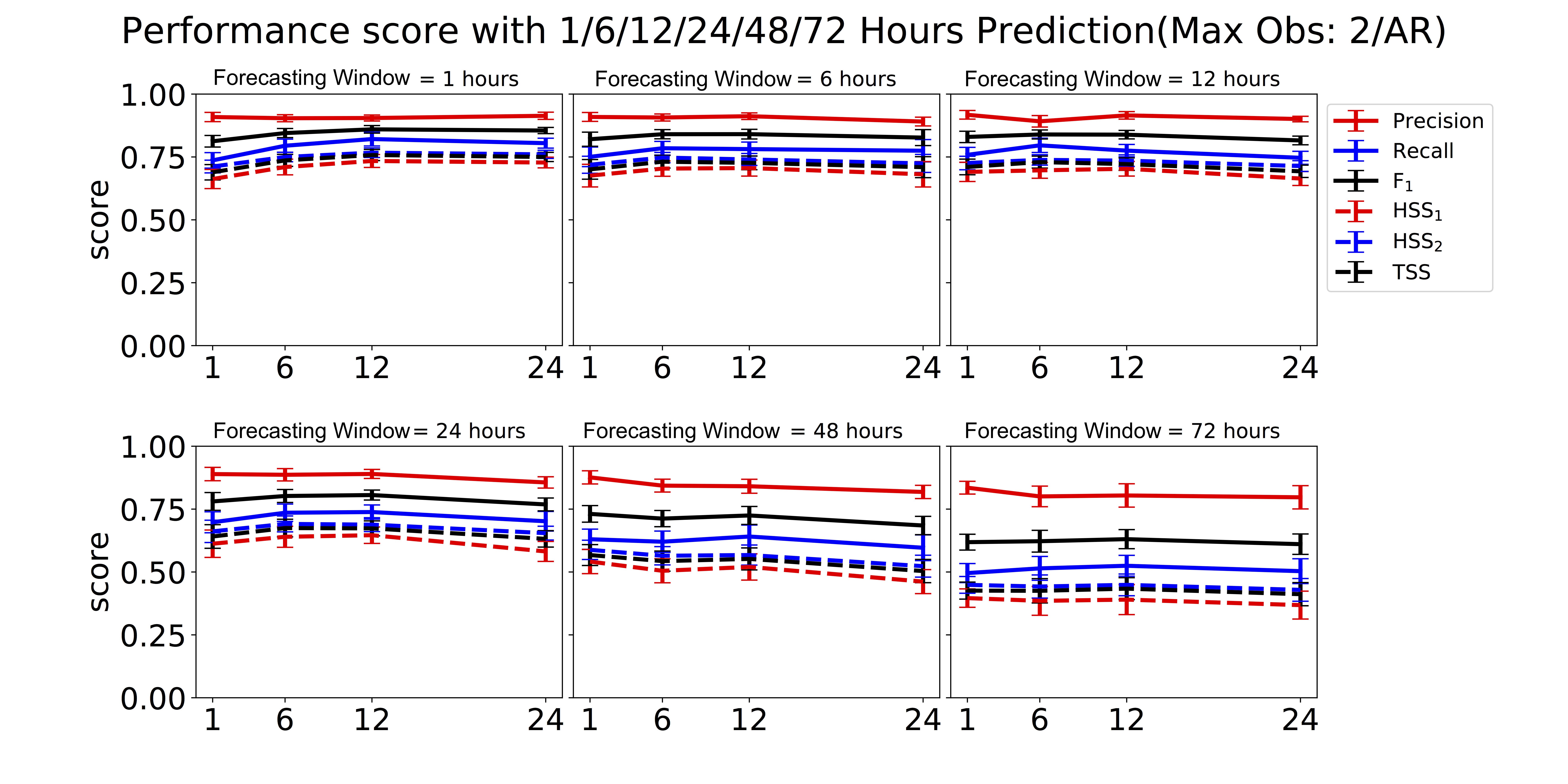}
    \caption{Performance scores from split-by-active-regions (with cap = 2, i.e. the number of events per AR is less than or equal to 2), as described in Section~\ref{subsec:data_split}, are displayed in the same way as in \figurename~\ref{fig:sw_f1} in Section~\ref{subsec:strong_weak_classification} in the main text.}
    \label{fig:split_ar_cap2}
\end{figure}

\begin{figure}[htb]
\centering
    \includegraphics[width=0.9\textwidth]{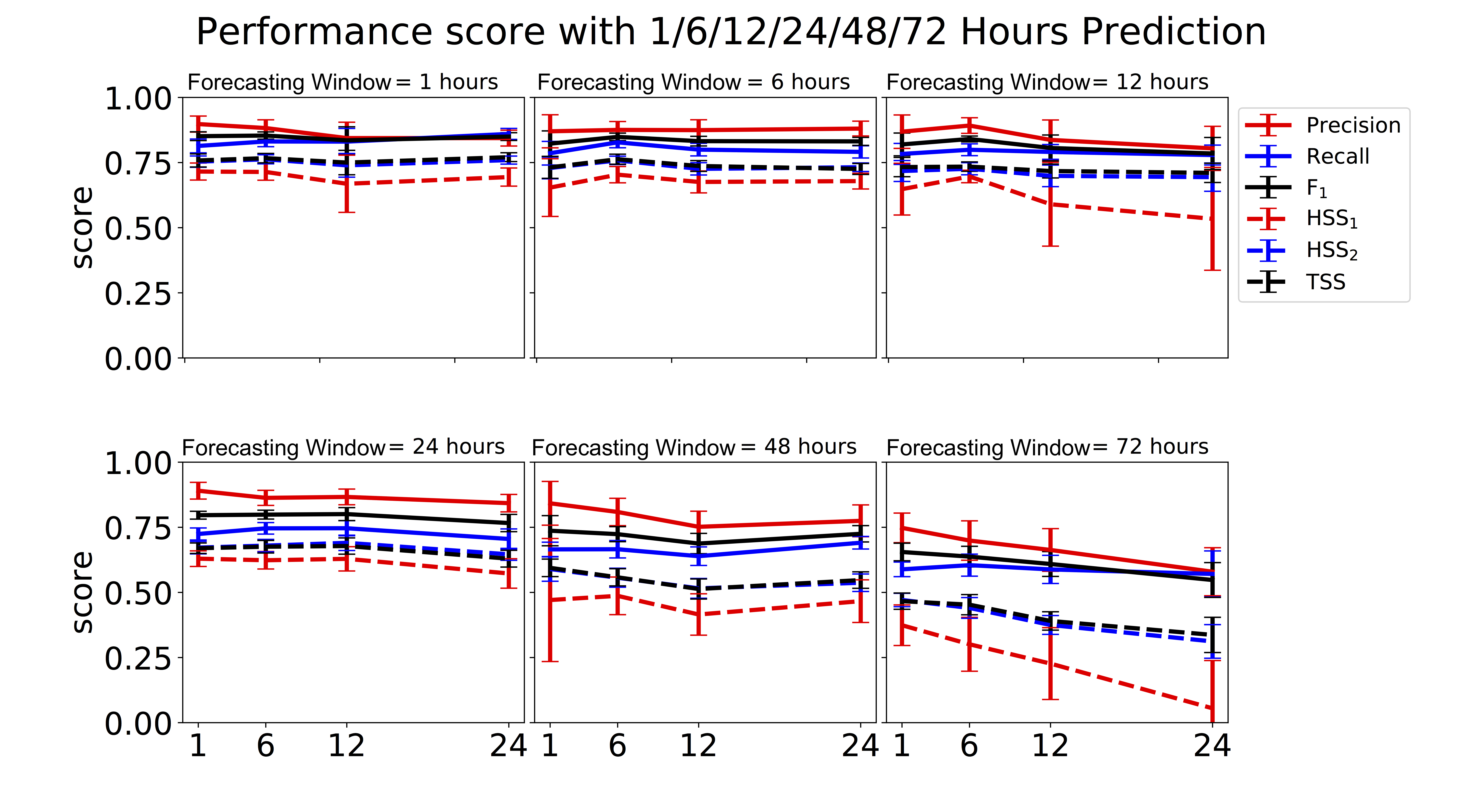}
    \caption{Performance scores from split-by-year randomly, as described in Section~\ref{subsec:data_split}, are displayed in the same way as in \figurename~\ref{fig:sw_f1} in Section~\ref{subsec:strong_weak_classification} in the main text.}
    \label{fig:split_year_climb}
\end{figure}

\begin{figure}[hb!]
    \centering
    \includegraphics[width=0.9\textwidth]{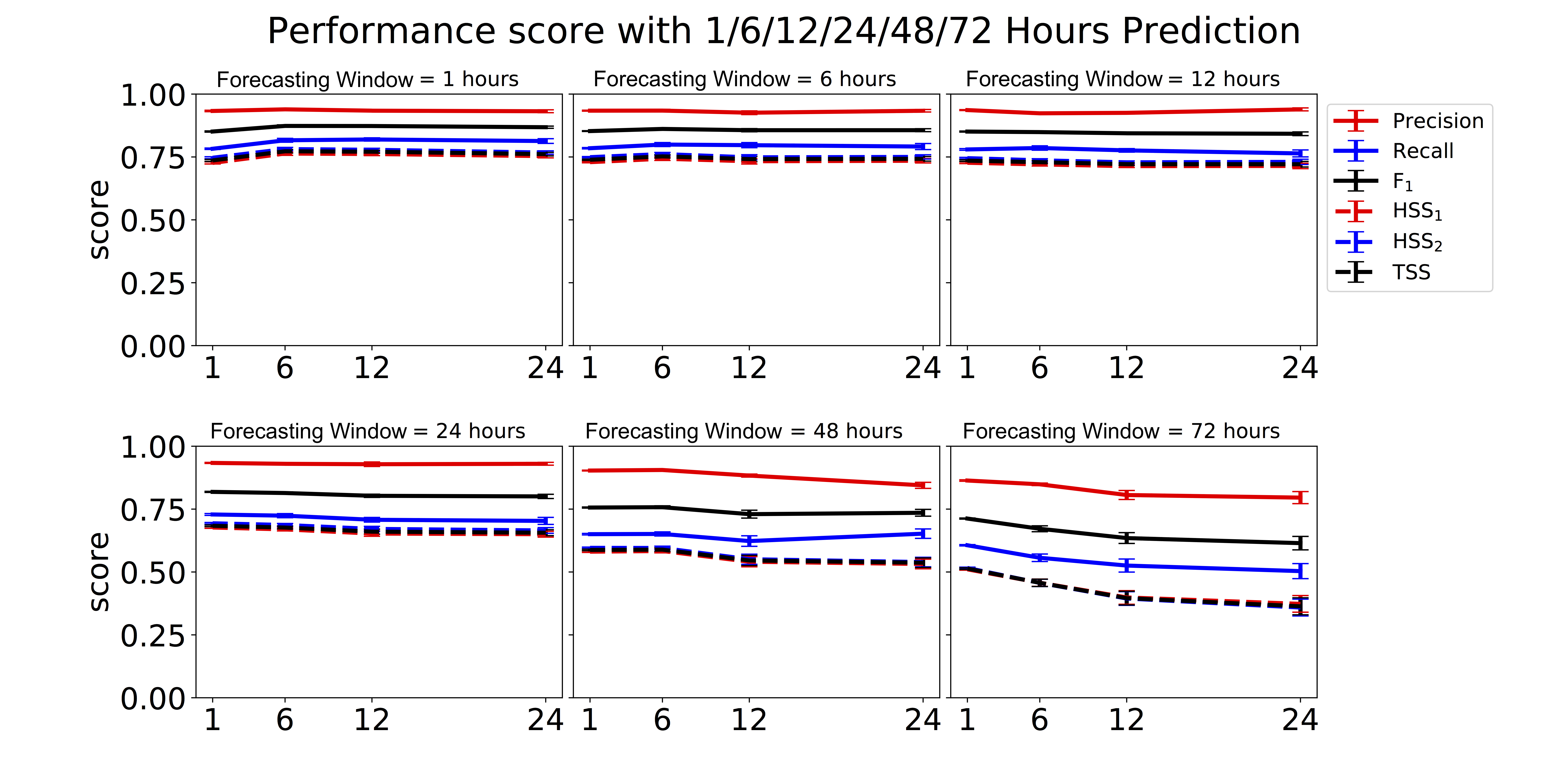}
    \caption{Performance scores from split-by-year (training with solar climbing and maximum and testing with solar declining phase), as described in Section~\ref{subsec:data_split}, are displayed in the same way as in \figurename~\ref{fig:sw_f1} in Section~\ref{subsec:strong_weak_classification} in the main text.}
    \label{fig:split_year_declining}
\end{figure}

The positive and negative classes are not balanced for the training and testing data when we put caps on the number of flare events per AR. We give the proportion of the positive class in the training \& testing data for all values of caps that we test in \tablename~\ref{tab:train_test_prop_split_AR}.

\end{appendices}



%
\newpage
\bibliographystyle{plainnat}
\bibliography{references_solar_flare.bib}

\end{document}